\documentclass[12pt,preprint]{aastex}

\slugcomment{accepted by ICARUS}

\shorttitle{Spectropolarimetry of Jupiter and Saturn}
\shortauthors{Schmid et al.}

\begin{document}


\title{Long slit Spectropolarimetry of Jupiter and Saturn\thanks{Based on 
observations obtained at the ESO 3.6m Telescope at La Silla, 
Chile (ESO program 72.C-0498)}}


\author{H.M.~Schmid, F.~Joos, E.~Buenzli and D.~Gisler}
\affil{Institut f\"ur Astronomie, ETH Z\"urich, 8093-Z\"urich, Switzerland
\footnote{E-mail addresses: schmid@astro.phys.ethz.ch (H.M. Schmid), 
ebuenzli@astro.phys.ethz.ch (E. Buenzli), 
gisler@astro.phys.ethz.ch (D. Gisler).}}

\begin{centering}
\vskip5cm
\noindent
41 pages, 11 figures, 3 tables \\
January 5, 2011 \\
\vskip1cm

running head: H.M. Schmid et al. Spectropolarimetry of Jupiter and Saturn\\
\vskip3cm

\noindent
address for proofs and correspondence: \\
H.M. Schmid \\
Institute of Astronomy \\
ETH Zurich  \\
Wolfgang Pauli Str. 27 \\
CH - 8093 Zurich \\
Switzerland \\
\medskip

\noindent
schmid@astro.phys.ethz.ch \\

\end{centering}

\newpage

\noindent
ABSTRACT
\medskip

\noindent
We present ground-based limb polarization measurements of Jupiter 
and Saturn consisting of full disk imaging polarimetry for the 
wavelength 7300~\AA \ and spatially resolved (long slit) 
spectropolarimetry covering the wavelength range 5200 to 9350~\AA. 

For the polar region of Jupiter we find for $\lambda=6000$~\AA \ a very 
strong radial (perpendicular to the limb) fractional polarization 
with a seeing corrected maximum of about $+11.5~\%$ in
the South and $+10.0~\%$ in the North. This indicates that the polarizing
haze layer is thicker at the South pole. The polar haze layers extend 
down to 58$^\circ$ in latitude. The derived polarization values are
much higher than reported in previous studies because of the better 
spatial resolution of our data and an appropriate consideration of the
atmospheric seeing. Model calculations demonstrate that the
high limb polarization can be explained by strongly polarizing 
($p\approx 1.0$), 
high albedo ($\omega \approx 0.98$) haze particles with a scattering 
asymmetry parameter of $g\approx 0.6$ as expected for aggregate particles
of the type described by \citet{west91}. 
The deduced particle parameters are distinctively different 
when compared to lower latitude regions. 

The spectropolarimetry of Jupiter shows  
a decrease in the polar limb polarization towards longer 
wavelengths and a significantly enhanced polarization in 
strong methane bands when compared to the adjacent continuum.
This is a natural outcome for a highly polarizing haze layer above
an atmosphere where multiple scatterings are suppressed in absorption
bands. For lower latitudes the 
fractional polarization is small, negative, and it depends only little 
on wavelength except for 
the strong CH$_4$-band at 8870~\AA .  

The South pole of Saturn
shows a lower polarization ($p\approx 1.0-1.5~\%$) than the poles of Jupiter.
The spectropolarimetric signal for Saturn decrease rapidly with
wavelength and shows no significant enhancements in the fractional 
polarization in the absorption bands. These properties can be 
explained by a vertically extended stratospheric haze region
composed of small particles $<100$~nm as suggested
previously by \citet{karkoschka05}.

In addition we find in the V- and R-band a previously not observed 
strong polarization feature ($p=1.5-2.0~\%$) near the equator of Saturn.  
The origin of this polarization signal is unclear 
but it could be related to a seasonal effect. 

Finally we discuss the potential of 
ground-based limb polarization 
measurements for the investigation of the scattering particles in the
atmospheres of Jupiter and Saturn. 


\newpage

\section{Introduction}
\label{sectintro}

Reflected light from planets is polarized. Polarimetry therefore provides 
information on the nature
and distribution of the scattering particles in the atmospheres of
planets complementary to other techniques \citep[see][]{coffeen79}. 
As the scattering polarization 
from extra-solar planets can be quite high 
\citep{kattawar71,seager00,stam04,buenzli09}, polarimetry is also used
for the search of extra-solar planets with existing and
future instruments \citep{hough06,schmid06b}
The polarimetric investigations of extra-solar planets also generate 
renewed interest in the polarization properties of solar
system planets. These can be used to predict the expected
polarization signal for extra-solar
planets, and thus help in the interpretation of future
detections. 

Solar system planets have been frequently observed polarimetrically until
1990 with instruments using single channel (aperture) detectors 
\citep[e.g.][]{leroy00}. However, almost no data were taken with
``modern'', ground-based imaging polarimeters and spectropolarimeters 
using array detectors. Therefore the polarimetric properties of 
solar system planets are still not well characterized.  

We have therefore started a program of ``modern'',
ground-based polarimetric
observations of solar system planets. In \citet{schmid06a} and 
\citet{joos07a} 
we described the data for Uranus and Neptune, for which we detected a
strong limb polarization. 
In this paper we present imaging polarimetry for Jupiter and Saturn, taken
with the Zurich imaging polarimeter (ZIMPOL) at the McMath-Pierce
solar telescope, and long slit spectropolarimetry taken with the EFOSC2 
instrument attached to the ESO 3.6m telescope. 

For the outer planets the possible phase angles for ground based 
observations are very limited, and the disk integrated polarization is
close to zero due to the back-scattering situation. But with spatially
resolved observations one can use the limb polarization 
effect to constrain polarimetric properties of the atmosphere. 
The limb polarization is a well-known second order effect 
occurring in reflecting
atmospheres where Rayleigh-type scattering processes are dominant
\citep[e.g.][]{vandehulst80}.
To understand this effect, one has to consider a back-scattering
situation at the limb of a sphere, where we have locally a configuration 
of grazing incidence and grazing emergence 
for the incoming and the back-scattered photons, respectively.
Photons reflected after one scattering
are unpolarized, because the scattering angle is 180$^\circ$. 
Photons undergoing two scatterings travel after the first scattering
predominantly parallel to the surface before being reflected towards 
us by the second scattering process. Photons going up will mostly escape
without a second scattering, and photons going down have a low 
probability of being reflected towards us after the second 
scattering but a high probability to be absorbed or to 
undergo multiple scatterings. Because the polarization angle induced 
in a single dipole-type scattering process, like Rayleigh scattering, 
is perpendicular to the propagation direction of the
incoming photon (which is often parallel to the limb), 
a net polarization perpendicular to the limb is produced.

Polarimetric data of Jupiter and Saturn were first published 
in the pioneering paper of \citet{lyot29}, 
who detected for Jupiter a strong positive polarization of 
$p\approx 5-8~\%$ at the poles with an orientation perpendicular 
to the limb. In the disk center he measured a phase angle dependent 
polarization which is essentially zero near opposition and slightly negative 
(parallel to the scattering plane), $p\approx -0.4~\%$, 
for phase angles around 
$\alpha=10^\circ$. These measurements were confirmed and 
improved in many later observations using single
aperture polarimeters \citep[e.g.][]{dollfus57,gehrels69,morozhenko73,hall76} 
and some imaging polarimetry by \citet{carlson89}.  

Important results on the polarization of Jupiter 
were achieved with the Pioneer 10 and 11 spacecrafts, which 
obtained polarization maps for phase angles larger than 
$\alpha= 12^\circ$. The data show that the polarization 
in the B- and R-band for $\alpha \approx 90^\circ$ reaches a level of
about $p\approx 50~\%$ at the poles, while the 
polarization is rather low ($< 10~\%$) in the equatorial region 
\citep[e.g.][]{smith84}. 
   
For Saturn, \citet{lyot29} found a 
phase dependent polarization for the rings and some
polarization for the atmosphere. 
Well established is the polarization of the ring system which shows 
for small phase angles $< 7^\circ$ a polarization of about
$p=-0.4~\%$, parallel to the 
scattering plane, with some dependency on the phase angle
\citep{johnson80,dollfus96}.
The disk of Saturn shows in the UV at 370~nm   
a strong radial limb polarization of 
more than $3~\%$ near opposition \citep{hall74}
as expected for Rayleigh scattering. In the visual the 
polarization is lower ($\lesssim 1~\%$), and predominately in N-S 
direction. The visual polarization shows significant temporal
variations which may be seasonal. The poles show usually but not
always the highest polarization 
\citep[e.g.][]{kemp73,hall68,dollfus96,gisler03}.

Polarimetry of Saturn for large phase angles $\approx 30^\circ-150^\circ$
was made with the spacecrafts Voyager 2 \citep{west83} and
Pioneer 11 \citep{tomasko84}. 
These data show a strong wavelength dependence of the
polarization with a low polarization ($<5~\%$) in the red, 
roughly $\approx 20~\%$ in the blue and $>30~\%$ (phase angle
$\alpha = 68^\circ$) in the UV at 264~nm. A big step forward is
expected from the imaging polarimetry of Saturn taken with
the Cassini spacecraft. \citet{west09} provide a 
first glimpse on a high quality Cassini polarization map
of Saturn taken in 2003 for a phase angle of 61$^\circ$. 

This paper presents imaging polarimetry and long
slit spectropolarimetry of Jupiter and Saturn. 
In the next section a description of our observations and the data
reduction are given. In Section 3 the polarization for Jupiter 
is described and analyzed while Saturn is treated in Section 4. 
In Section 5 the observations are compared with model simulations and
a qualitative interpretation is given for some polarization features.  
The results are discussed in the final section.

\section{Observations and data reduction}

Observations of Jupiter and Saturn were
taken in November 2003 in spectropolarimetric mode with the ESO 3.6m telescope
at La Silla, and in March 2002 and March 2003 with polarimetric imaging 
at the McMath-Pierce solar telescope at Kitt Peak. 
Observational parameters for Jupiter and 
Saturn were taken or derived from data given in \citet{almanach03} 
and they are summarized in Table \ref{parameter}. 

The illumination of the planet and the scattering geometry for the
reflected light is defined by the phase angle $\alpha$, which
is the angle sun-planet-observer, and the position angle (PA) of the
scattering plane $\theta$ with respect to the central meridian (North-South
direction) of the planet. The bright limb is on 
the East for $\theta$ close to 
$90^\circ$ and on the West for $\theta$ close to $270^\circ$.


The angles $\alpha$ and $\theta$ are important parameters 
for polarimetry. The strength of polarimetric features depends 
on the phase angle $\alpha$, and the orientation of the induced
scattering polarization is often
perpendicular or parallel to the orientation of the scattering plane
$\theta$. The scattering plane is essentially in East-West direction 
for $\theta$ close to $90^\circ$ or $270^\circ$.
For the November 2003 observation of Saturn, the scattering plane 
is tilted by 
$+12^\circ$ with respect to the East-West direction. In this case 
a perpendicular or parallel polarization with respect 
to the scattering plane will produce besides a $Q$-polarization 
(in N-S orientation)
also a significant $U$-polarization component of 
$U=0.45\,Q$. For a tilt angle of $\Delta\theta=- 2^\circ$ (e.g. 
Nov. 2003 for Jupiter) this factor is $ U =- 0.07\, Q$.  

The apparent diameters $d_{\rm N-S}$ and $d_{\rm E-W}$ are used to 
convert locations $x$ from the disk center (= sub-earth point) along the
central meridian (CM) to radial distances 
$r_{\rm CM}=x\cdot d_{\rm N-S}/2$ which can be
converted to planetographic latitudes 
considering the ellipsoidal shape and 
the inclination of the planets (Table \ref{parameter}).  

\begin{table}[h]
\centering
\caption{Parameters for the Jupiter and Saturn observations. $\alpha$
  is the phase angle, $\theta$ the orientation of the scattering plane, and 
  $r_{\rm CM}$ are distances from the sub-earth point on 
  the central meridian.}
\begin{tabular}{lcccc}
\hline
parameter & Jupiter && Saturn \cr
\hline
date (2003)      & March 9    & Nov. 29      & March 8     & Nov. 29       \cr
observatory      & Kitt Peak   & La Silla     & Kitt Peak    & La
Silla      \cr
instrument       & ZIMPOL      & EFOSC2       & ZIMPOL       & EFOSC2     \cr
$\alpha$         & $6.9^\circ$ & $10.4^\circ$ & $6.3^\circ$  & $3.7^\circ$   \cr
$\theta$ 
                 & $269^\circ$ & $88^\circ$   & $272^\circ$ & $102^\circ$  \cr
polar axis incl. & $+0.2^\circ$ & $-1.5^\circ$ & $+27.0^\circ$ & $+25.0^\circ$ \cr
lat. sub-earth point  & $+0.2^\circ$ & $-1.6^\circ$ & $-32.0^\circ$ & $-29.9^\circ$\cr
diameter (E-W)   & $43.57''$   & $35.91''$    & $18.63''$    & $20.23''$    \cr
$r_{\rm CM}$ limbs        
                 & $\pm 20.38''$& $\pm 16.79''$ & $\pm 8.59''$ & $\pm 9.30''$    \cr
$r_{\rm CM}$ south pole   
                 & $-20.37''$  & $-16.78''$   & $-7.49''$    & $-8.27''$    \cr
$r_{\rm CM}$ equator      
                 & $+0.1''$    & $-0.5''$     & $+4.2''$     & $+4.3''$   \cr
$r_{\rm CM}$ ring inner edge N
                 &             &              & $+6.45''$    & $+6.46''$ \cr
$r_{\rm CM}$ ring outer edges 
                 &             &              & $\pm 9.59''$ & $\pm 9.61''$ \cr
\hline
\end{tabular}
\label{parameter}
\end{table}

\subsection{Spectropolarimetry}
\label{sectspecpol}
Spectropolarimetric observations of Jupiter and Saturn were taken 
during the nights of November 29 and 30, 2003 with EFOSC2 at the ESO
3.6m telescope at La Silla. These data originate from the same run 
and instrument setup as the spectropolarimetry of Uranus and Neptune 
from \citet{joos07a,joos07b}, where descriptions of the
measuring strategy and the data reduction are given. Here we provide 
only a brief outline and highlight some special points. 

EFOSC2 at the Cassegrain focus of the ESO 3.6m telescope is a multi-mode
imager and  grism spectrograph which can be equipped with a 
Wollaston prism and a rotatable super-achromatic half-wave plate for 
linear polarimetry and 
spectropolarimetry. A special slit mask can be placed in the focal
plane which consists of a series of 19.7$''$ long slitlets with a
period of 42.2$''$  appropriate to avoid overlapping of the spectra from
the ordinary and extraordinary beam of the Wollaston prism. The width
of the slitlets used for Jupiter and Saturn was 0.5$''$.

For extended sources the EFOSC2 instrument setup provides long slit
spectropolarimetry and thus the intensity and polarization as function
of wavelength $\lambda$ and position $x$ along the slit. 
The orientation of the slit can be changed by rotating the whole instrument. 
The spatial resolution or the effective seeing 
of our data is about 1$''$,
as derived from the width of the spectra of the standard stars.  
 
Most of the Jupiter and Saturn data were taken with the slitlets along
the central meridian. Some data were taken in East-West orientation 
(coordinates of the planet), but due to instrument flexure 
problems the calibration of the E-W slit data turned out to be less accurate
(see below). 

The grism (EFOSC2 grism\#5) employed provided the wavelength range from
5200 to 9350\,\AA\ with a resolution of 6.4\,\AA\ for a 0.5$''$ wide
slit. The data were recorded with a $2{\rm k} \times 2{\rm k}$ CCD
(ESO CCD\#40) with a spatial scale of $0.157''$ and a spectral scale
of 2.06\,\AA\ per pixel. For $\lambda>7000$\,\AA\ the CCD introduces 
an interference
pattern. To smooth this pattern and to enhance the signal-to-noise
ratio of the data the spectra are binned along the wavelength
direction into 30\,\AA\ bins. 

The linear polarization was measured in a standard way 
\citep[e.g.][]{tinbergen92}, with sets of four exposures taken with
half-wave plate positions at $0^\circ$, $22.5^\circ$, $45^\circ$ and
$67.5^\circ$ respectively. For better data quality several sets 
were taken for each slit orientation. The exposure time
per frame was 3~s for Jupiter, and 5~s for
Saturn. Polarimetric standard stars were observed with the same
instrumental setup to determine the polarization offset introduced
by the telescope and to obtain the polarization angle zero point. The
orientation of the slit for standard star observations was always in  
celestial north-south direction. Exposures of a helium-argon lamp
provided the wavelength calibration.

The data reduction was performed with the {\sc midas} software
package. For long slit spectropolarimetry it is important that the
spectra of the ordinary and extraordinary beam are aligned with
an accuracy of about 1/10 of a pixel in spatial direction, to ensure
that no artificial polarization is introduced. This precision was
achieved for the observations with the slit in N-S direction (N-S for
the planets was not far from N-S on the sky) using the
standard star data as reference \citep[see][]{joos07a}.
The calibration of data taken with the slit in E-W direction 
suffered from instrument flectures and 
the standard stars spectra could not be used as alignment reference. 
Therefore the E-W spectra were
aligned with respect to the edge of the slitlets. The achieved
alignment precision is only about half as good as with the help of the
standard stars. 

The instrumental polarization of EFOSC2 at the Cassegrain focus of the
3.6m telescope is low, and the polarimetric calibration of the data
is straightforward. For the calibration we used the unpolarized standard
star HD 14069 and the polarized standard star BD $+25^\circ 727$. 
The instrumental polarization was found to be less 
than 0.2\,\% in the central region of the field. 
The polarization angle calibration should be accurate to about
$\Delta\theta\approx \pm 2^\circ$. 

No flux calibration was attempted. Solar and telluric spectral 
features in the intensity spectra were corrected
with the help of Mars observations taken with the same instrument
configuration. The overall slope of the intensity spectrum was
adjusted to the albedo spectrum from \citet{karkoschka98}. 
 
The observations of the linear polarization for Jupiter 
and Saturn are given as Stokes parameters $I(\lambda,x)$, 
$Q(\lambda,x)$, $U(\lambda,x)$, where $I$ is the total
intensity and $Q=I_0-I_{90}$, $U=I_{45}-I_{-45}$ are 
the polarized fluxes and $Q/I$, $U/I$ the
fractional (normalized) Stokes parameters. We also use the term 
radial polarization,
e.g. $Q_r$ or $Q_r/I$, indicating a polarization
parallel (positive) or perpendicular (negative) to the slit  
through the disk center. 
The orientation of $+U_r$ is rotated by $45\degr$ in counter-clock
direction (North over East).  

\subsection{ZIMPOL imaging polarimetry}
\label{ObsZimpol}

Imaging polarimetry of Jupiter and Saturn was taken in March 2002 and 2003 at
the McMath-Pierce solar telescope on Kitt Peak observatory, using
ZIMPOL the Zurich Imaging Polarimeter \citep{povel95,gisler04}.

ZIMPOL is a high precision polarimeter, consisting of a fast 
polarization modulator, a polarizer, and a camera with a 
special masked CCD sensor. The polarization 
state of the incoming light is changed into a temporal polarization variation 
by the modulator, which is subsequently converted into an intensity variation 
by the linear polarizer. The masked CCD has periodically arranged 
open and covered rows. During integration the photo-charges 
are shifted back and forth between the open and one or more covered 
rows synchronously with the modulator. In this way the CCD camera 
has two or more image planes for the different polarization modes 
(e.g. $I_0$, $I_{90}$ ...), and all these images are registered 
with the same CCD pixels. Flat-fielding
problems, differential aberrations, and alignment errors 
are minimized with this technique and the polarization can be
determined with high sensitivity.

This paper is focussed on the analysis of the spectropolarimetric signal
of the strong polarization features of Jupiter and Saturn. The ZIMPOL imaging
polarimetry is treated as useful auxiliary data because they 
cover the whole planetary disk of Jupiter and Saturn 
and Saturn's ring system in one
exposure. This helps to verify and complete the spatial dependence of the
spectropolarimetric data, for which one slit setting provides only
incomplete spatial coverage.   

We have selected for our analysis the ZIMPOL imaging 
polarimetry of Jupiter and Saturn 
taken in March 2003 in a filter with a central wavelength of 7300~\AA\ 
and a band width of 200~\AA. Observational 
parameters for these data are given in Table
\ref{parameter}. The pixel scale was 0.41 arcsec/pixel and 
the seeing was near 1.5 arcsec.
In addition we use for Jupiter a ZIMPOL intensity image in the
6010~\AA \  ``continuum''
filter (width 180~\AA )
for the calibration of the relative reflectivity scale on the disk. 

A drawback of the ZIMPOL / McMath-Pierce telescope 
observations are the position and wavelength dependent 
instrument polarization effects
introduced by the inclined mirrors of the solar telescope. 
We modelled the telescope polarization, but uncertainties
at a level of $\Delta
Q/I, \Delta U/I\approx \pm 0.5~\%$ remain.

\begin{figure}
\plottwo{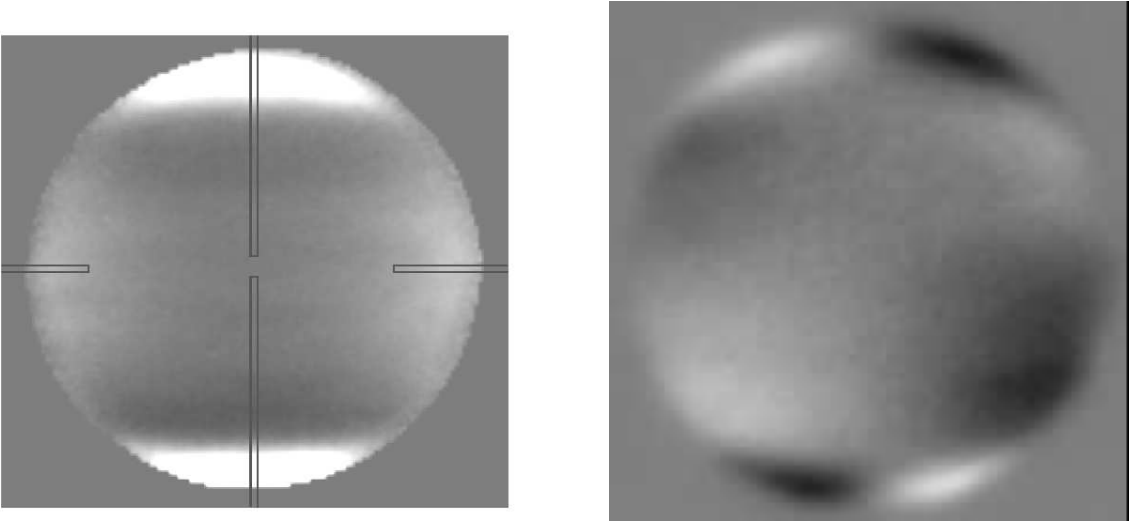}{jspolfig1b.eps}
\caption{$Q$ (left) and $U$ (right) polarization flux images of
 Jupiter in a filter centered at 7300\,\AA\
  taken with ZIMPOL in March 2003 at Kitt Peak. North is up and East 
  is left. The grey scale is normalized to the central intensity and
  spans the range from $-1.0~\%$ (black) to $+1.0~\%$ (white). The
  lines in the $Q$-image indicate the slit positions for the 
  spectropolarimetric observations with EFOSC2.}
\label{jupdani}
\end{figure}

The instrument polarization can be corrected based on 
a priori assumption for the instrument and the targets.
First we assumed that the 
instrumental polarization is field independent. Second we used the known
polarization properties of Jupiter and Saturn. The reflected 
light shows no circular polarization at a level 
$|V/I| <0.1~\%$ \citep{kemp71,swedlund72,smith83}. Further, the existing 
maps show an $U/I$ polarization which 
is essentially anti-symmetric with respect to the central 
meridian and a disk integrated $U$-polarization close to 
zero $|U/I|<0.1~\%$ \citep[e.g.][]{hall68,hall74}.
This should also hold for the
ZIMPOL data, because they were taken for an epoch where the scattering
plane for Jupiter and Saturn was essentially in East-West direction;
With these assumptions we can derive from our full Stokes polarimetry 
a cross talk corrected $Q/I$-image (see Fig.~\ref{jupdani}).
This $Q/I$-image still includes 
all instrument polarization offsets induced by the inclined telescope
mirrors. This offset was derived for our observations of Jupiter 
from the polarization phase curves of \citet{morozhenko73}, 
using the value $Q/I=-0.2~\%$ for the disk center with an 
uncertainty of less than $\pm 0.1~\%$. For Saturn the ring polarization 
is reasonably well known \citep[see e.g.][]{dollfus96,johnson80}
and for the offset correction we adopt a value of $Q/I=-0.4~\%$ for 
the polarization of the East and West cusps.    


\noindent

\section{Jupiter}

\subsection{Imaging polarimetry}

Figure \ref{jupdani} shows the ZIMPOL Stokes-$Q$ and $U$ images of Jupiter
taken in March 2003 in the 7300~\AA\ filter. A strong positive $Q$ or $U$
polarization flux is plotted white, while black indicates a strong
negative polarization, and grey denotes little or no polarization flux. 

The $Q$ flux image clearly shows the two strongly 
polarized poles with a polarization in N-S direction. 
Further there is a weak N-S polarization
(parallel to the limb) near the equatorial
limbs. In the center of the disk the polarization is slightly negative
in East-West direction.
The $U$ image shows at the poles 
a positive or negative component on the East and
West side of the central meridian. This indicates that the
polarization at the poles is not in N-S but in radial
direction.

The fractional $Q/I$ polarization at the poles reaches values 
of about $+7~\%$ in the 7300~\AA \ filter. The disk-integrated 
(flux weighted) $Q/I$-polarization is only $+0.2~\%$. The polarization
of the disk center is slightly negative according to the
``imposed polarization'' offset calibration 
in accordance with the polarization phase curves of
\citet{morozhenko73}. Our spectropolarimetric data
(Sect.~\ref{Jspecpol}) confirm the calibration of the imaging
polarimetry.

\subsection{Limb to limb profiles}
 
\subsubsection{Comparison of North-South and East-West profiles}

\begin{figure}
\epsscale{0.8}
\plotone{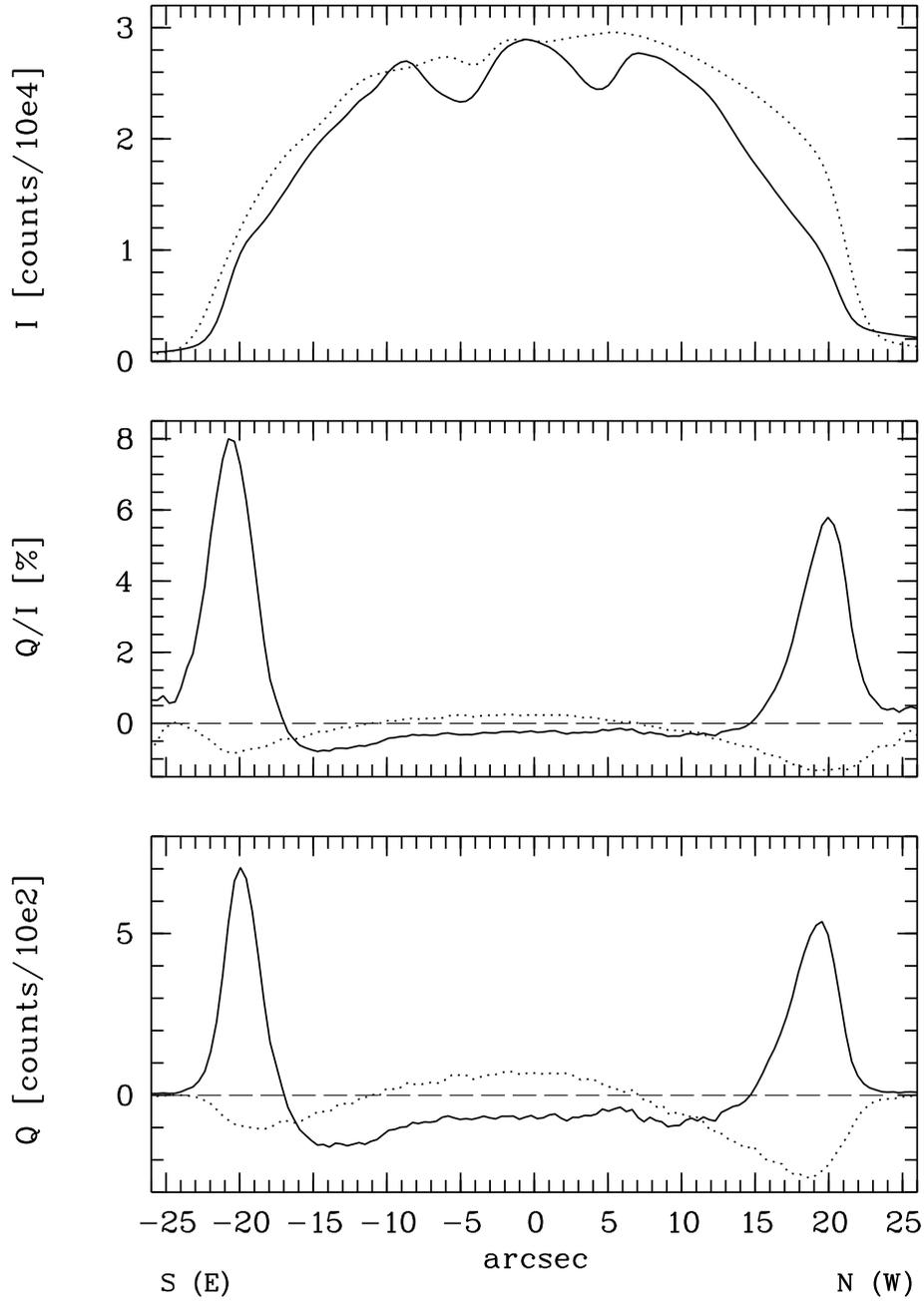}
\caption{Observations of Jupiter from March 2003. 
  Polarimetric profiles in N-S (full line) and E-W
  (dotted line) direction, taken in the 7300~\AA\ filter
  with ZIMPOL. The top panel shows the intensity $I$, the middle panel the
  fractional (radial) polarization $Q_r/I$, and the bottom panel the
  polarization flux $Q_r$.}
\label{danicut}
\end{figure}

Figure \ref{danicut} shows the ZIMPOL 7300~\AA \  intensity 
and polarization profiles
through the disk of Jupiter in North-South direction 
along the central meridian
and in East-West direction along the equator. 
The plots give the radial polarization where 
positive $Q_r$ or $Q_r/I$ values stand for a polarization
perpendicular to the limb. 
The position is indicated in arcsec with $0''$ at the center of the
apparent disk. The nominal limb position is at $\pm 20.4''$ for the
N-S profile and at $\pm 21.8''$ for the 
E-W profile (see Table \ref{parameter}). 

The N-S intensity profile in the 7300~\AA\ filter shows Jupiter's 
dark bands and bright zones structure very similar to many previous
studies \citep[e.g.][]{chanover96,moreno91,west79}.
The main feature of the equatorial intensity profile 
(dotted line) is the asymmetry due to the $\alpha=6.9^\circ$ 
illumination offset, with the bright limb in the West.   
 
The N-S and E-W fractional polarization profiles $Q_r/I$ 
illustrate the huge differences between the poles with
a very high, positive limb polarization 
$Q_r/I \approx 8~\%$ in the South and $\approx 6~\%$ in 
the North and the equatorial region with a small
negative polarization. 
The polarization in the disk center is parallel to the scattering
plane, which translates into 
a radial polarization $Q_r$ with a negative sign for the North-South
profile and a positive sign for the East-West profile. 

The presented N-S polarization profile for 7300~\AA \ agrees well with 
earlier measurements for the visual-red spectral region, e.g. 
from \citet{lyot29}, \citet{dollfus57}, 
or \citet{hall68}. The polarization in the disk 
center depends 
on phase and it varies from $Q/I\approx 0.0~\%$ for $\alpha=0^\circ$ 
to about $Q/I\approx -0.5~\%$ 
for $\alpha=12^\circ$ as described in detail in 
\citet{morozhenko73}. 
A negative limb polarization for the equator region was also previously
reported for small phase angles $\lesssim 6^\circ$ and
the red spectral region \citep{dollfus57,gehrels69}.
In the UV/blue spectral region 
the limb polarization at the equator is due to Rayleigh scattering
perpendicular to the limb \citep{hall68,gehrels69}.
 
The radial polarization in $U_r$-direction is 5 to 10 times weaker and the 
signal obtained is dominated by systematic noise. The $U_r$-signal and 
the uncertainties are
quantified for the spectropolarimetric observation.

\subsubsection{Polarization measurements for the central meridian}
\label{JupMeridian}

The long-slit spectropolarimetric measurements provide
polarimetrically well calibrated profiles for the central meridian for all 
wavelengths between 5300 and 9300\,\AA . 

\begin{figure}
\epsscale{0.8}
\plotone{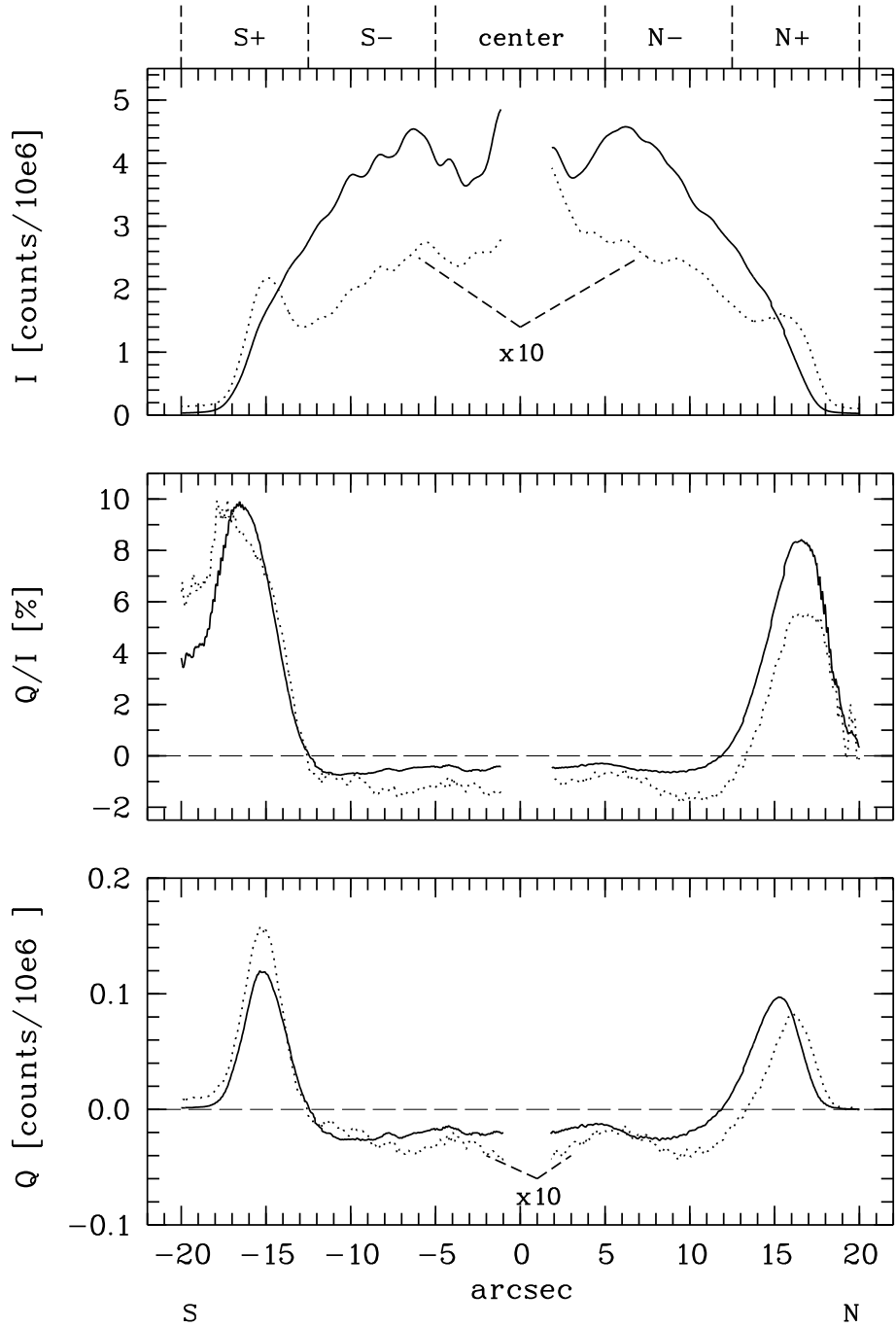}
\caption{Observations of Jupiter from November 2003. 
  Intensity $I$, fractional polarization $Q_r/I$, and
  polarization flux $Q_r$ profiles through the central meridian (N-S) 
  of Jupiter for the
  continuum at 6000~\AA\ (full line) and the deep methane absorption band
  at 8870\,\AA\ (dotted line). The $I$ and
  $Q_r$ flux for the methane band are multiplied by 10 for visibility
  reasons. At the top the five selected spatial regions are indicated.}
\label{jupNScut}
\end{figure}

Figure \ref{jupNScut} shows profiles for 
the continuum at 6000~\AA , spectrally averaged from 5900 to
6100\,\AA , (solid line) and the deep methane absorption band at
8870\,\AA\, spectrally averaged from 8800 to 8940\,\AA. Spectral
averaging was done to enhance the signal to noise.

The intensity profiles show belt and zones at 6000~\AA\,  and a 
significant limb brightening in the 8870~\AA\ methane band 
in good agreement with previous studies 
\citep[e.g.][]{moreno91,chanover96}.     

The $Q_r/I$ and $Q_r$ 
polarization profiles from November 2003 in Fig.~\ref{jupNScut} 
are very similar to the profile from March 2003 shown in
Fig.~\ref{danicut} with $Q_r/I\approx 10$~\%
at the south pole for both spectral bands. 
At the north pole $Q_r/I$ for 
6000~\AA\, reaches a maximum value just above 8~\%, while the
maximum for 8870~\AA\, is slightly less than 6~\%. 
It is interesting to note that the South pole shows not only a
stronger polarization than the North pole, but also a stronger limb
brightening in the methane absorption $\lambda$8870. 
The polarisation is negative in the center of the disk. The
sign change occurs at about $\pm 12.5$ arcsec, corresponding to a
Jovian latitude of about $\pm 59^\circ$. 

It is not trivial to quantify accurately the measured polarization signal.
The peak polarization at the limb depends significantly on the spatial
resolution (or the effective seeing) and it is difficult to compare
our measurements with previous data which often depend on the spatial
resolution of the measurement. 

In order to provide quantitative results we split the profile into the
polar sections with positive polarization S+ and N+ and sections with negative
polarization S$-$ and N$-$. The section borders are set at $\pm 12.5$ arcsec,
where the signs change, and at $\pm 5$
arcsec, the approximate end of the slitlets. For measurements of the
central region without spectropolarimetric coverage we interpolated
the curves based on the ZIMPOL $\lambda$7300 imaging 
polarimetry and intensity profiles from the
literature \citep[e.g.][]{chanover96}.   
 
Table \ref{jupitertable} gives the flux $I$, polarization fluxes $Q_r$
and $U_r$, and the flux weighted fractional 
polarization $Q_r/I$ and $U_r/I$ for the wavelengths 
6000~\AA, 8200~\AA, and 8870~\AA. The flux is given as ratio
$I/I_{\rm slit}$ relative to the flux for the entire slit. 
This quantity does not depend on uncertainties 
in the absolute flux calibration.  

\begin{table}
\caption{Fractional intensity $I/I_{\rm slit}$, 
  radial polarization $Q_r/I$ and $U_r/I$ for
  Jupiter for the sections S+, S$-$, center, N$-$, N+ along the 
  central meridian with boundaries as defined in the first two
  rows. $A_g$ is the geometric albedo for the considered wavelength range
  from \citet{karkoschka98} and $f$ the derived reflectivity
  (see text).} 
\begin{tabular}{lrrrrrr}
\noalign{\smallskip\hrule\smallskip}
                 & S+   & S$-$   & center & N$-$ & N+  & total slit \cr
\noalign{\smallskip\hrule\smallskip}
\noalign{location}
$x_{\rm min}$    & $-20''$ & $-12.5''$ & $-5''$ & $+5''$ & $+12.5''$ & $-20''$
                 \cr
$x_{\rm max}$    & $-12.5''$ & $-5''$ & $+5''$ & $+12.5''$ & $+20''$ & $+20''$ 
\cr
$\Delta x/x_{\rm slit}$   
                & 0.128$*$ & 0.223 & 0.298 & 0.223 & 0.128$*$ & 1.000$*$ \cr
\noalign{continuum 5900-6100 \AA , $A_g=0.54$ \hfill}
$I/I_{\rm slit}$ & 0.067 & 0.250 & 0.366 & 0.250 & 0.067 & 1.000 \cr
f            & 0.31     & 0.66   & 0.72    & 0.66    & 0.31  & 0.59   \cr
$Q_r/I$ [\%] & $+4.36$ & $-0.56$ & $-0.44$ & $-0.44$ & $+3.95$ & $+0.14$ \cr
$U_r/I$ [\%] & $+0.35$ & $0.00$  & $-0.01$ & $-0.05$ & $-0.31$ & $-0.02$ \cr
$Q_r/I_{\rm slit} [\%]$
             & $+0.292$ &        &         &         & $+0.265$  \cr
\noalign{continuum 8100-8300 \AA , $A_g=0.48$ \hfill}
$I/I_{\rm slit}$ & 0.057 & 0.246 & 0.368 & 0.255 & 0.074 & 1.000 \cr
f            & 0.23     & 0.57   & 0.64    & 0.59    & 0.30 & 0.52     \cr
$Q_r/I$ [\%] & $+3.31$ & $-0.68$ & $-0.42$ & $-0.61$ & $+1.73$ & $-0.16$ \cr
$U_r/I$ [\%] & $-0.40$ & $-0.02$ & $+0.01$ & $+0.02$ & $-0.24$ & $+0.01$ \cr
$Q_r/I_{\rm slit} [\%]$
             & $+0.189$ &        &         &         & $+0.128$  \cr
\noalign{CH$_4$-band 8800-8940 \AA , $A_g=0.05$ \hfill}
$I/I_{\rm slit}$ & 0.095 & 0.203 & 0.388 & 0.229 & 0.096 & 1.000 \cr
f            & 0.039    & 0.047   & 0.068   & 0.053   & 0.039 & 0.052 \cr
$Q_r/I$ [\%] & $+5.69$ & $-1.16$ & $-1.07$ & $-1.21$ & $+2.79$ & $-0.15$ \cr
$U_r/I$ [\%] & $+0.61$ & $+0.07$ & $+0.26$ & $+0.02$ & $-0.63$ & $+0.10$ \cr
$Q_r/I_{\rm slit} [\%]$
             & $+0.541$ &         &         &         & $+0.268$  \cr
\noalign{\smallskip\hrule\smallskip}
\end{tabular}
$*$: $\Delta x/x_{\rm slit}$ considers only the slit section 
located within the limb at $\pm 16.79''$.  \\
\label{jupitertable}
\end{table}

The values $Q_r/I$ and $U_r/I$ for the S+
and N+ regions in Table~\ref{jupitertable} depend strongly on the exact
size of the integration interval due to the steep intensity slope 
$I(x)$ at $x=\pm 12.5''$. Much less critical is the determination 
of the total polarization 
flux $Q_r$ from $x=\pm 12.5''$ (where $Q_r\approx 0$) to the
limb. The measured value can be related to $I_{\rm slit}$, giving
$Q_r(S+)/I_{\rm slit}$ and $Q_r(N+)/I_{\rm slit}$
according to $Q_r/I_{\rm slit} = I/I_{\rm slit} \cdot Q_r/I$
(also for $U_r/I_{\rm slit}$). This relative limb polarization flux
is an interesting quantity for long
term variability studies, since it does not depend much on the spatial
resolution of the observations and can be derived from the 
counts data without flux calibration or conversion 
into normalized reflectivity. It remains to be determined whether
$Q_r(S+)/I_{\rm slit}$ and $Q_r(N+)/I_{\rm slit}$ depend on phase
angle.

A conversion of the fractional intensity  
$I/I_{\rm slit}$ into reflectivity $f$ can be made.  
If the average reflectivity along the central meridian  
$\langle f_{\rm slit} \rangle$ is known, then the reflectivity in a bin
is 
\begin{displaymath}
f=\langle f_{\rm slit} \rangle {I/I_{\rm slit} \over x/x_{\rm
    slit}}\,.
\end{displaymath}
From the full disk image we can derive the ratio 
$\Lambda = \langle f_{\rm disk} \rangle /\langle f_{\rm slit} \rangle$
between the average reflectivity for the 
full planetary disk ( = the geometric albedo $A_g$) and $f_{\rm slit}$.
With the geometric albedo from the literature
one can determine
\begin{displaymath} 
\langle f_{\rm slit} \rangle =  {A_g\over \Lambda}\,.  
\end{displaymath}
We derive $\Lambda=0.92$ from the ZIMPOL
image for the ``continuum'' filter centered 
at 6010~\AA \ (width 180~\AA). This value is not much
different from $\Lambda_{\rm Lam} = 0.85$ for a perfectly 
white Lambert sphere.  
\citet{karkoschka98} gives 
for this wavelength a geometric albedo of $A_g=0.59$ for data
taken in 1995. This value can be used for our calibration because 
the global reflectivity variations
of Jupiter are small ($\la 5~\%$) and also the phase dependence 
of the reflectivity ($\approx 1.5~\%$ for $\alpha \approx 0^\circ-10^\circ$)
can be neglected. 
$\Lambda=0.92$ can also be employed to derive the reflectivities
for the continuum wavelength
8200~\AA , because our intensity profiles along
the central meridian have a very similar shape for wavelengths from
5300~\AA \ to 8700~\AA . 

More difficult is the calibration of the reflectivity for the strong
methane band at $\lambda=8870~$\AA. This profile shows 
a relatively high reflectivity at the equator and the poles and
relatively dark mid-latitudes (see Fig.~\ref{jupNScut}). The overall 
reflectivity profile is flatter than for the continuum wavelengths 
$6010$~\AA \ or $8200$~\AA , but not completely flat ($\Lambda=1$) as for
a reflecting disk. Strongly absorbing atmospheres, like Jupiter in a
strong CH$_4$ band, have a rather constant reflectivity over the
disk \citep[see][]{buenzli09}. Based on these considerations we 
adopt $\Lambda=0.96$ for the 8870~\AA -methane band.   

The uncertainties for the fractional
polarization values $Q_r/I$ and $U_r/I$ given in 
Table \ref{jupitertable} are mainly due to systematic
errors like instrument calibration or inaccuracies in the slit
positioning. 
%
%
The derived $Q_r/I$ and $U_r/I$ values should be accurate to 
$\Delta(Q_r/I), \Delta(U_r/I)\approx \pm 0.1~\%$ to $0.2~\%$, except for 
the $U_r/I$-values for S+ and N+, for which a $Q\rightarrow U$
cross-talk error or/and slit positioning error at a level of up to $\Delta
U_r/I\approx \pm 0.3~\%$ to $\pm 0.6~\%$ could be present. We therefore
conclude that our data show no significant $U_r$ 
component for the central meridian of Jupiter.            

\subsection{Spectropolarimetry}
\label{Jspecpol}

\begin{figure}
\epsscale{0.8}
\plotone{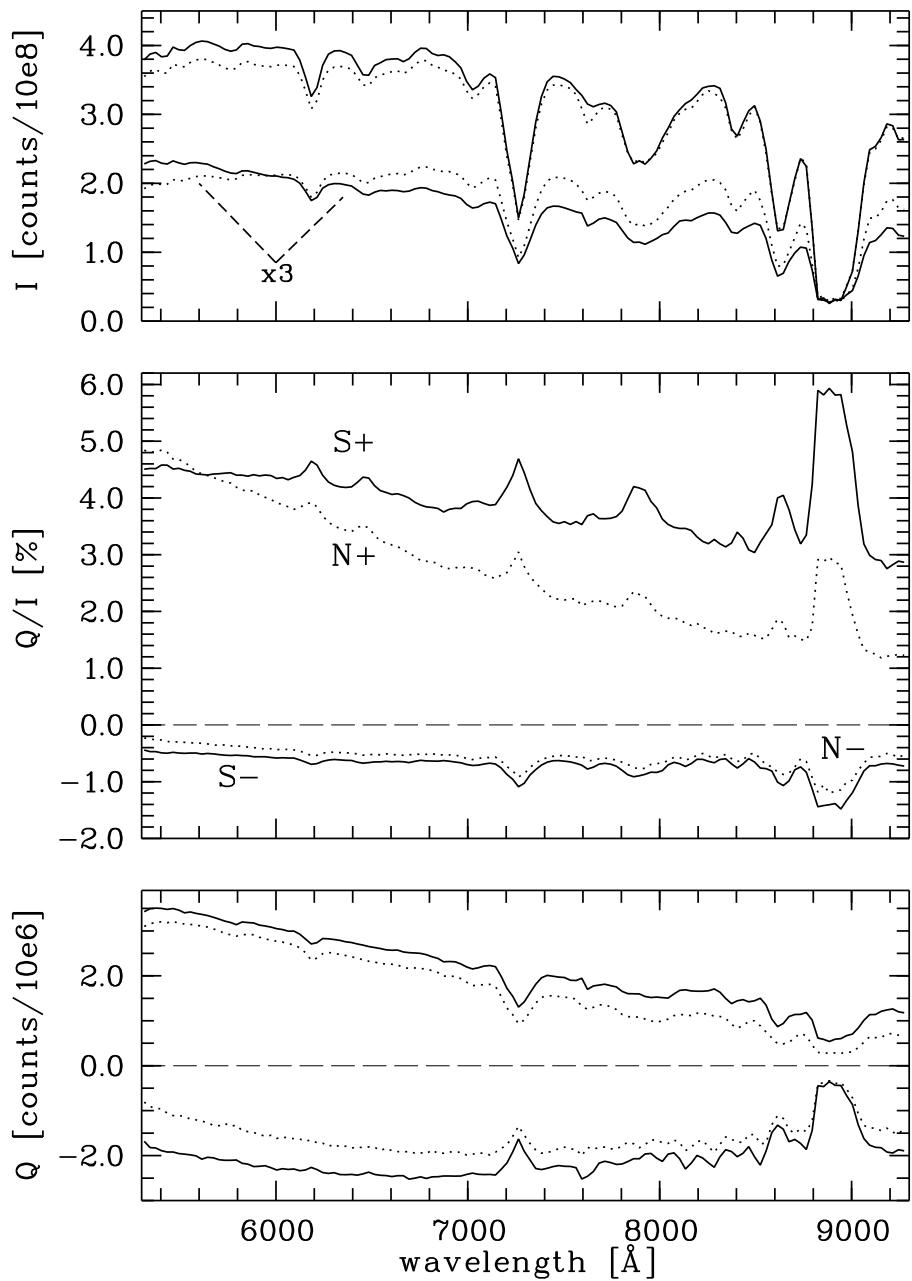}
\caption{Spectropolarimetry of Jupiter for the S+, S$-$ (solid lines)
  and the N$-$ and N+ (dashed lines) regions as defined in
  Fig.~\ref{jupNScut}. The intensity $I(\lambda)$ for the
  polar region S$+$ and N$+$ are multiplied by a factor of 3 with
  respect to S$-$ and N$-$ for visibility reasons. The middle panel
  gives $Q_r/I(\lambda)$ and the bottom panel the polarization flux $Q_r$.}
\label{jupspec}
\end{figure}

The EFOSC2 observations provide the spectropolarimetric signal for each 
point along the slit. A general picture is provided in 
Fig.~\ref{jupspec} which shows averages for the 
spatial regions S+, S$-$, N$-$, and N+, defined 
in the previous section. 

The reflection spectra $I(\lambda)$ are color-calibrated with respect to 
the full disk albedo spectrum of 
\citet{karkoschka98}. Spectroscopic features are very similar 
for all regions \citep[see also][]{cochran81} but systematic 
differences do exist. For example the equivalent width of the strong
$\lambda 8870$ CH$_4$ band is smaller for the limb regions S+ and N+,
when compared to mid latitudes S$-$ or N$-$. This is just another
manifestation of the limb brightening effect for CH$_4$ $\lambda$8870 
seen in Fig.~\ref{jupNScut}.

The fractional polarization $Q_r/I(\lambda)$ shows for the South 
polar region S+ a decrease of the continuum polarization with wavelength 
from about $Q_r/I=4.6~\%$ at 5300~\AA\ to 2.8~\% at 9300~\AA . The decrease is
much steeper for the northern limb from $Q_r/I=5.0~\%$ to 1.2~\%. 
In the strong absorption bands $Q_r/I$ is 
enhanced with respect to the adjacent continuum, most prominently in
the CH$_4$-band at $\lambda 8870$.  
 
The fractional polarization for the mid-latitude regions S$-$
and N$-$ is low, but negative, and the continuum polarization increases
slightly from about 
$Q_r/I=-0.3~\%$ at 5300~\AA\ to $-0.6~\%$ at
9300~\AA. The polarization in the methane bands is also enhanced, in
this case more negative 
(Fig.~\ref{jupspec}, Table \ref{jupitertable}). 

The polarization flux spectrum $Q_r(\lambda)=I \cdot Q_r/I$ shows a 
reduced signal at the position of the strong absorption bands. 
This indicates that the enhancement of $Q_r/I$ in
the absorption bands is smaller than the reduction in $I$, so that 
the polarization flux spectra still exhibit a
reduction in $|Q(\lambda)|$ at the wavelengths of the
absorptions.  

Essentially no previous spectropolarimetric data from Jupiter is
available in the literature apart from the multi-filter aperture 
polarimetry of \citet{gehrels69} and the detection of
a weak differential polarimetric signal due to a methane band by
\citet{smith83} measured with
full disk observations.
\citet{gehrels69} found a similar wavelength 
dependence for the polar limb polarization. 
Specifically, in 1960/63 they measured for 
longer wavelength ($>5000$~\AA) a higher polarization at
the northern than at the southern limb  -- the opposite N-S 
asymmetry when compared to our observations. 
They also found that the polarization
asymmetry is reversed for shorter wavelength ($<5000~$\AA), and the
South pole shows a higher polarization. Such a reversal of the
N-S polarization asymmetry also seems to be present at the
short wavelength end of our data.

\subsubsection{Spectropolarimetry for the ``extreme'' polar limbs}
 
\begin{figure}
\includegraphics[angle=-90,scale=.50]{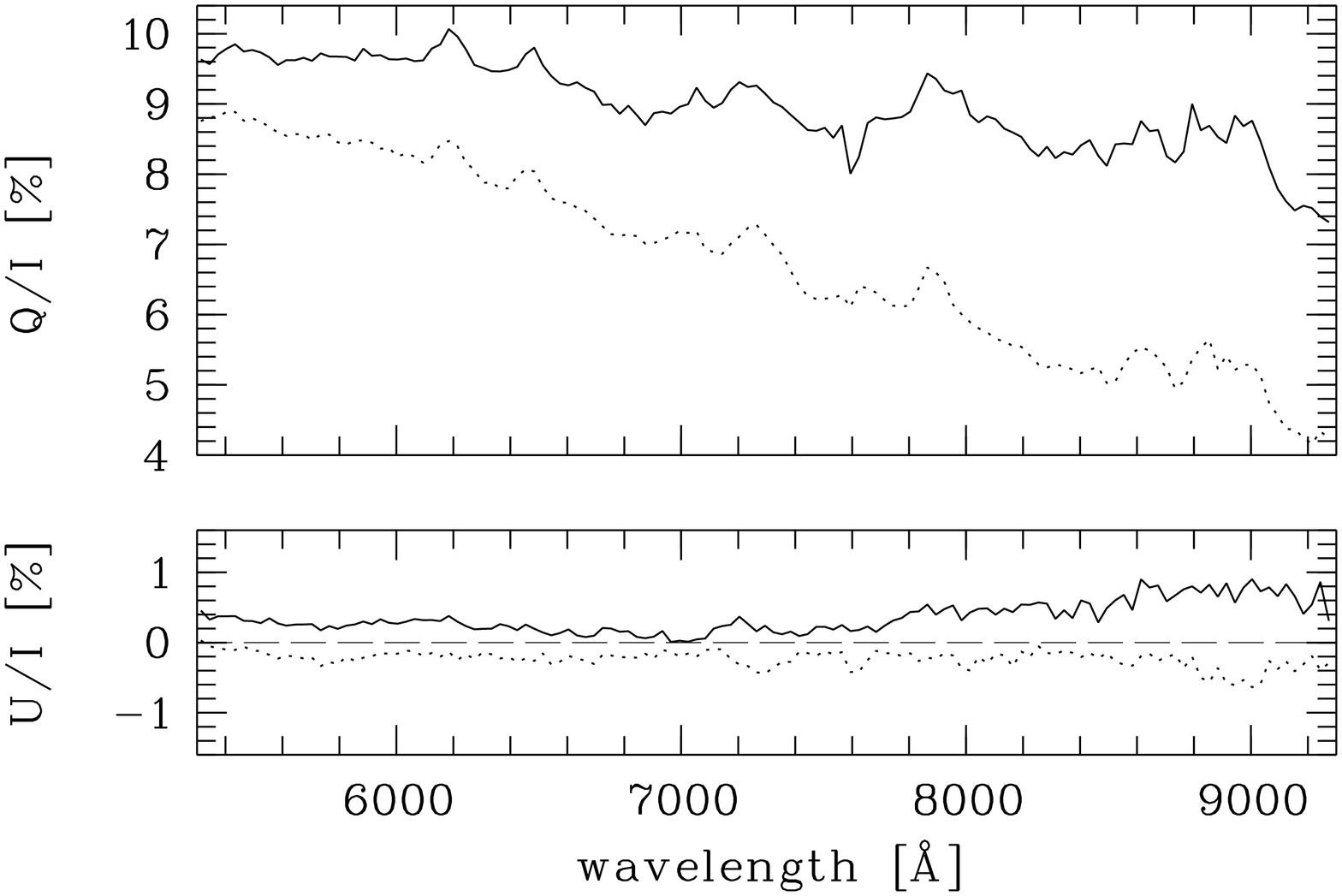}
\caption{Fractional polarization spectra $Q_r/I$ and $U_r/I$ 
for the ``extreme'' southern (solid) and the northern (dashed) limb.}
\label{jupspecmax}
\end{figure}

The S+ and N+ spectropolarimetry provides only a very coarse
characterization of the ``average'' limb polarization. For this reason we
explore in this section the spectropolarimetric signal at the
``extreme'' polar limbs, where we find the maximum of the fractional 
polarization $Q_r/I$. We select a small spatial bin from 
$\pm 16''$ to $\pm 17''$, corresponding to planetographic latitudes from
about $\pm 73^\circ$ out to the polar limbs (see Fig.~\ref{jupspecmax}). 

For the South pole the fractional polarization reaches 
$Q_r/I=9.6~\%$ for the continuum
in the 5300 to 6100~\AA\ region and even 10.0~\% in the 
$\lambda$6190 CH$_4$-band. Previous studies report maximum 
fractional polarizations of not more than $Q_r/I=7-8~\%$, 
most likely because the spatial resolution was worse than the $1''$ achieved 
with our observations. 
This illustrates the problem of the spatial averaging. 
The overall spectral slope of the fractional 
polarization spectra $Q_r/I(\lambda)$
at the extreme polar limb is flatter for the South than for the North,
similar to the S+ and N+ regions. Interestingly the 
polarization enhancement in the strong methane bands is 
much less pronounced in the
extreme polar limb data.

\subsubsection{Spectropolarimetric signal at the equatorial limb}

The spectropolarimetric observations of Jupiter taken with a slit 
in East-West direction cover the regions from the eastern limb at 
$-18.0''$ to $-12.8''$ and from $+10.7''$ 
to the western limb at $+18.0''$ (Fig.~\ref{jupdani}). 

For both limbs the fractional polarization spectra show a
positive polarization $Q_r/I \approx 0.5~\%$ for the shortest 
wavelengths, and $Q_r/I \approx 0.0~\%\pm 0.2\%$
for $\lambda > 7000$~\AA , except for the strong CH$_4$ band at
8870~\AA , where the
polarization is approximately $Q_r/I \approx + 0.4$~\%. 

The spatial profiles show a general radial decrease in the 
fractional polarization from slightly positive values, 
$Q_r/I\approx +0.5~\%$ to $0.0~\%$,  
at the inner edges of the slitlets (at $-12.8''$ and $+10.7''$) towards zero
or slightly negative values, $Q_r/I \approx 0.0~\%$ to $-0.5~\%$, further
out at the eastern limb and out to $+16''$ on the western side. 
For a given distance from the center, the polarization is 
slightly more negative ($\Delta p\approx - 0.2~\%$) 
on the eastern side. 

This behavior is qualitatively in agreement
with the E-W polarization profile derived for the ZIMPOL polarimetry
for March 2003. The E-W asymmetry is switched most likely 
because the bright limb and terminator have switched sides between
March 2003 and Nov. 2003. 
Also the small change in the overall polarization level for the phase
angle $\alpha=6.9^\circ$ in March 2003 (less E-W polarization in the 
equator region) and $\alpha=10.4^\circ$ in Nov 2003 (more E-W polarization) 
is similar to previous studies \citep[e.g.][]{morozhenko73}. 

\section{Saturn}

\subsection{Imaging polarimetry}
\begin{figure}
\epsscale{0.8}
\plotone{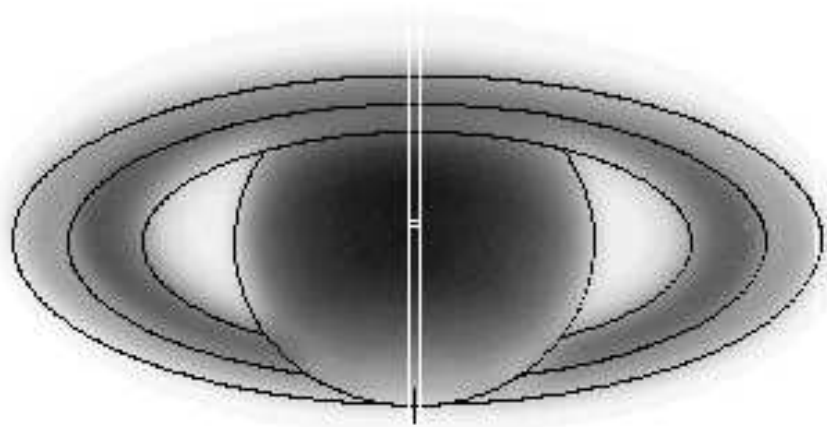}
\caption{EFOSC2 acquisition image of Saturn taken on Nov. 30, 2003. North is
 up and East is to the left. The limb of the planetary disk, as well as the 
 position of the south pole and the limits of the A and B rings are 
 indicated. In addition, the two slit positions for spectropolarimetry
 are marked with white rectangles which overlap at the center.}
\label{saturnpointing}
\end{figure}

The appearance of Saturn depends strongly on the inclination of
the planet and its ring system. In 2003 the inclination of the polar
axis with respect to the celestial plane was $27^\circ$,
close to the maximum inclination possible for Earth-bound
observations. An EFOSC2 acquisition image shows Saturn for the
November 2003 run (Fig.~\ref{saturnpointing}). For this 
inclination the South pole is well visible about $1''$ inward from the
limb (see Table \ref{parameter}). The equator is almost half way between
disk center and northern limb while the ring system covers the 
latitudes northward of $+20^\circ$. 

\begin{figure}
\centering
\includegraphics[angle=-90,scale=.45]{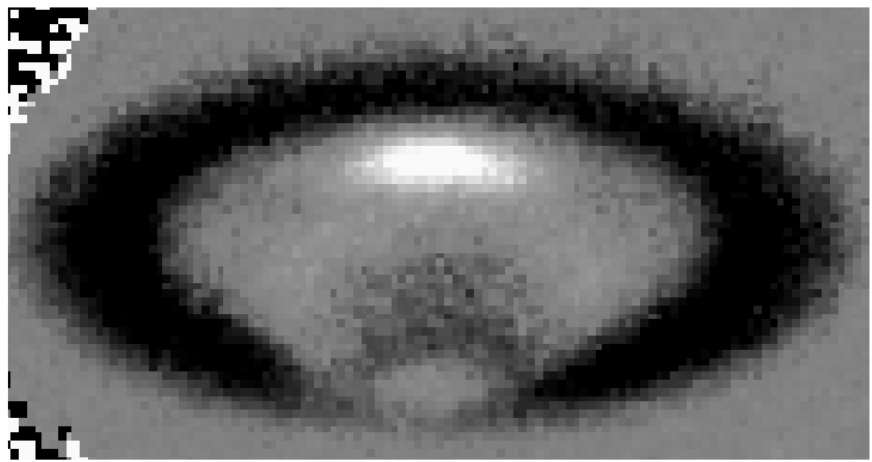}
\caption{ZIMPOL Stokes $Q$-flux image of Saturn, obtained in the 
methane 730~nm filter. In the south the planet occults the ring, except
for a small rim with a width of about $1.5''$. 
}
\label{saturnQ}
\end{figure}

Figure \ref{saturnQ} shows a Stokes $Q$-polarization flux image in the
7300~\AA\ filter taken with ZIMPOL in March 2003. One can
recognize the negative (E-W) polarization of the ring, 
a weak positive feature 
at the southern pole, a polarization flux close to zero in the south, 
corresponding to latitudes around $-60^\circ$, 
and a dominant strong positive feature at the equator. 

The $U$-image corresponding to Fig.~\ref{saturnQ} shows near the
southern limb 
negative $U/I\approx -0.3$~\% and positive $\approx
+0.3$~\% features on the east and west sides of the South pole, respectively 
very much
like the $U$-pattern for the poles of Jupiter. This indicates that
the polarization features at the South pole have 
a radial polarization direction. For the
rings of Saturn there is essentially no $U$-signal visible.

The equatorial polarization reaches in the 7300~\AA\ filter a maximum
fractional polarization of $Q/I \approx + 1.6$~\% and extends about
4$''$ to 5$''$ in east-west direction. This equatorial feature is remarkable 
since it was not present in early measurements 
\citep[e.g.][]{hall74,dollfus96}.
A similar signal is also present in the 6010~\AA\ imaging
polarimetry from the same date and    
in the 5500~\AA, 6010~\AA, and 7300~\AA\  filter ZIMPOL 
polarimetry from March 2002 \citep{gisler03}.
Also our spectropolarimetric observations of 
November 2003 confirms the presence of this transient signal.

The $U$-signal at the equator is weak with a fractional polarization
of $U/I\approx -0.3$~\% at the east limb,
$-0.1$~\% at the meridian, and about $+0.1$~\% on the
west limb. These are only small signals when compared
to the fractional $Q/I$-polarization of $\approx +1.5$~\% at the
intersection of meridian and equator. This indicates that
the equatorial polarization extends from about $-30^\circ$ to
$+30^\circ$ in longitude along the equatorial band and  
has a predominant orientation in north-south direction.

\subsection{Profiles for the planet and the ring}

\subsubsection{Comparison of North-South and East-West profiles}

\begin{figure}
\epsscale{0.8}
\plotone{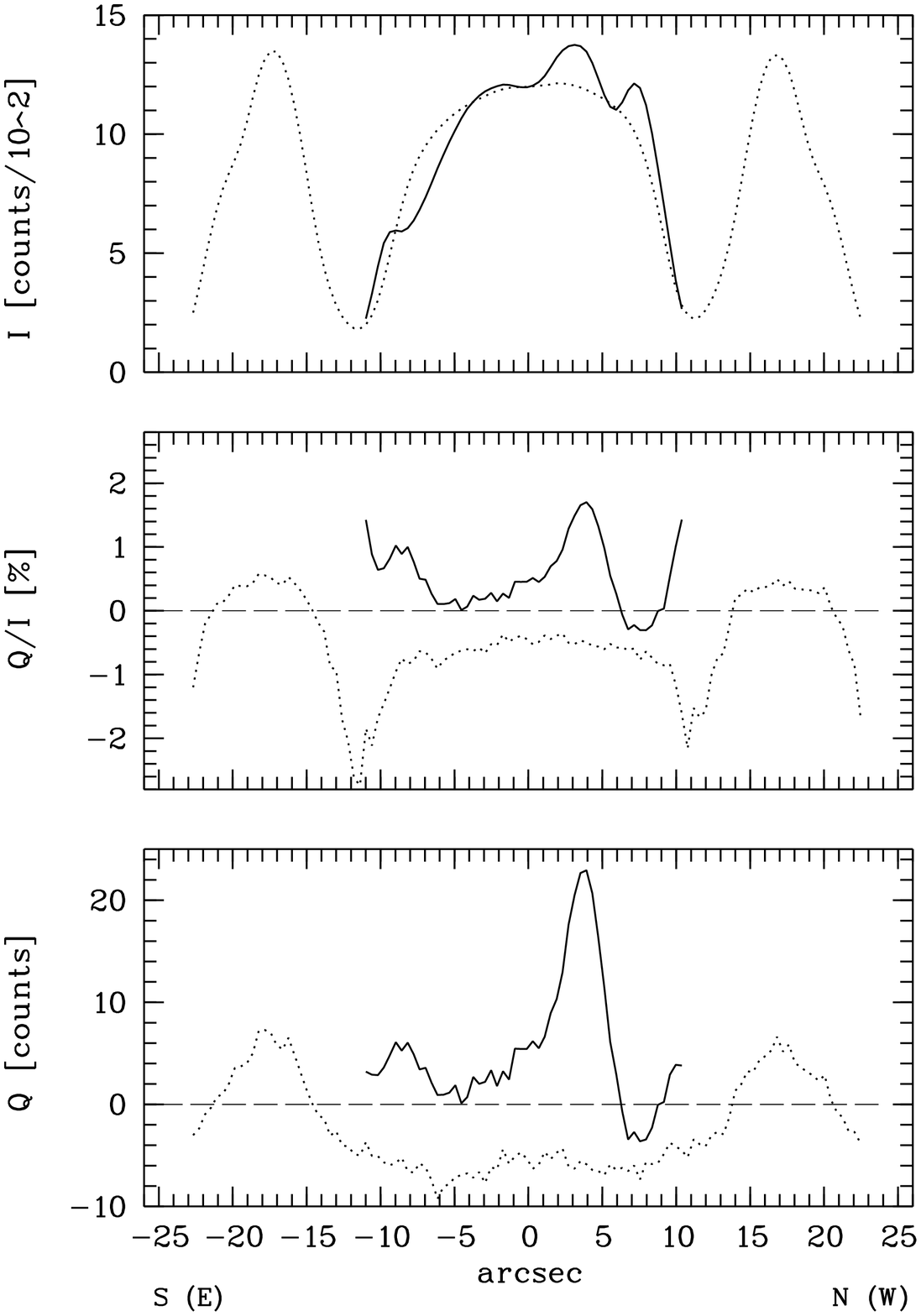}
\caption{Saturn N-S (full line) and E-W profiles (dotted line) 
for the intensity (top), the fractional radial polarization $Q_r/I$, 
and the $Q_r$ polarization flux 
for the wavelength $\lambda$7300\AA.}
\label{saturndanicuts}
\end{figure}

Figure \ref{saturndanicuts} shows the North-South and East-West 
profiles extracted for 
the 7300 \AA \ CH$_4$ filter ZIMPOL observations 
of March 2003. 
%
The middle panel shows the fractional radial polarization $Q_r/I$. 
The N-S profile for $Q_r/I$ is positive everywhere 
on the planetary disk and negative for the ring. There are two peaks, one of $Q_r/I\approx +1.0~\%$ 
at the southern pole (at approximately $-8.5''$) and a stronger peak of  
$Q_r/I\approx +1.6~\%$ at the equator ($+4''$) coincident with the 
intensity maximum.  
For the E-W direction the radial polarization is positive for 
the ring (parallel to the slit) and negative for the planet. The strongly 
negative $Q_r/I$ polarization 
between planet and ring is probably a spurious effect because in 
these gaps the photon
statistics is low and the impact of scattered light, and residual
systematic noise effects are large. 
The polarization flux profile $Q_r$ shows that the equatorial 
polarization feature is really strong.

As described in Sect.~\ref{ObsZimpol}
the calibration for the ZIMPOL-polarimetry is based on the
assumption that the fractional $Q/I$-polarization is $-0.4~\%$
for the ring. The good agreement between the ZIMPOL and
EFOSC2 data support this calibration procedure.  

\subsubsection{Polarization measurements for the central meridian}
\begin{figure}
\epsscale{0.8}
\plotone{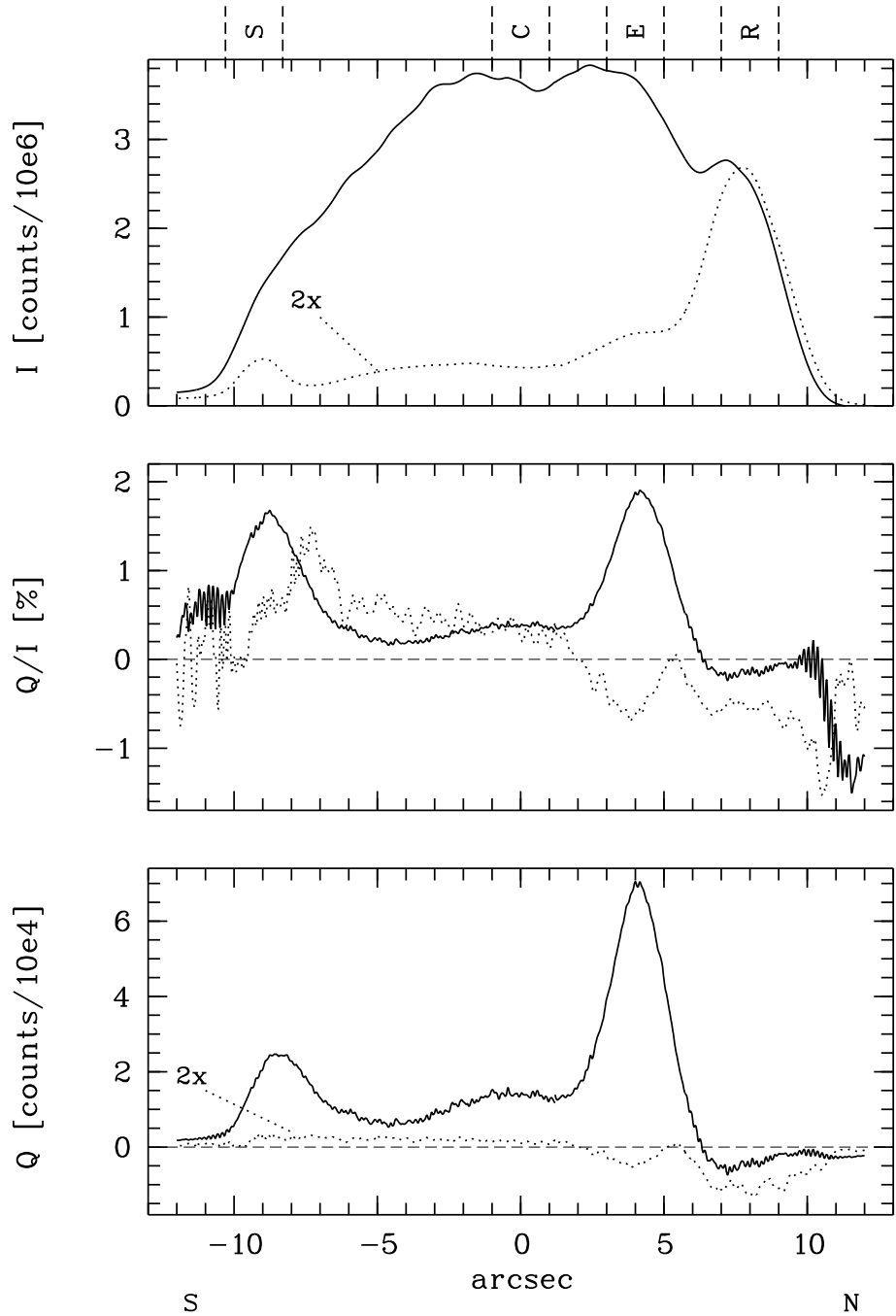}
\caption{Polarimetric profiles of Saturn along the central meridian 
  for the continuum wavelength region 6000\,\AA\ (full line) and the deep
  methane absorption band at 8870\,\AA\ (dotted line). 
} 
\label{saturncut}
\end{figure}

Figure \ref{saturncut} compares the meridional profiles 
for the strong methane band $\lambda$8870 and the continuum 
at $\lambda$6000. For the $\lambda$8870 CH$_4$ band the 
fractional polarization 
$Q_r/I$ is positive for the southern latitudes and negative at the
equator and the ring. It appears that $Q_r/I$ has a maximum at 
the southern pole at about $-7.5''$. 
The polarization flux $Q_r$ for 
$\lambda$8870 illustrates the weakness of the polarization signal
at this wavelength. In $Q_r$ no polarization flux maximum is
visible at the south pole. Interestingly there is a small 
negative polarization for the CH$_4$ wavelength near the equator 
in contrast to the strong positive polarization signal at shorter wavelengths.  

We select four regions along the the central meridian with a
width of $2''$ as indicated in Fig.~\ref{saturncut}. 
These regions are ``S'' for southern limb, 
``C'' for the center of the disk, ``E'' for the 
equator, and ``R'' for the ring and their
relative intensity 
and fractional polarization are given in 
Table \ref{saturntable}.

For the interpretation of the polarization in Table \ref{saturntable}
the orientation of the scattering plane has to be taken into account.
For the November 2003 observations the scattering plane is tilted by
$+12^\circ$ with respect to the E-W direction of Saturn. Therefore a scattering
polarization perpendicular to the scattering plane produces besides a
$Q$-signal also an $U$ signal at the level of $U=0.45\,Q$. 
For example, the $\lambda$6000 polarization for the ``S'' 
region has an orientation of $\theta_p=+5^\circ$, 
compatible with $\theta_p=0^\circ$ for a polarization in 
N-S direction (or radial), while the strong polarization   
in the equatorial region ``E'' has an orientation
of $\theta_p = +11^\circ$, perpendicular to $\theta=102^\circ$ of the
scattering plane. 

For the polarization of the E and W cusps of the ring  
the literature gives values of $\approx - 0.4~\%$ parallel 
to the scattering plane for phase angles near $4^\circ$ 
\citep[e.g.][]{dollfus79,dollfus96,johnson80}. For the ring ``R'' 
we measure for 
$\lambda$6000 and $\lambda$8100 a $Q/I$ polarization component 
between $-0.1~\%$ and $-0.2~\%$, or a polarization which is lower 
(less negative) by about 
0.2~\% than expected. This effect has been previously described 
by \citet{kemp73}, and at that time it was explained
by an admixture of $+Q/I$-polarized light from Saturn 
transmitted through the ring. Since we know now that the optical 
depth of the ring is rather high, we attribute this decrease in
polarization to light from the adjacent equatorial region scattered
by the instrument.
In the dark CH$_4$ band the reflection from the planet is strongly 
reduced, so that
the ring polarization is less diluted by scattered light from the planet. 

The uncertainties for the fractional
polarization due to instrument calibration and slit positioning errors
are about
%
$\Delta Q_r/I, \Delta U_r/I\approx \pm 0.2~\%$. Photon noise
errors are negligible compared to calibration errors and unidentified 
instrumental effects.

\begin{table}
\caption{Relative intensity $I/I_{\rm slit}$, reflectivity $f$, 
  and fractional radial polarization 
  $Q_r/I$ and $U_r/I$ for Saturn for the four $2''$ wide 
  regions S, C, E, R on the central meridian and the entire meridian. 
  The first row
  gives the center of these meridional sections and the second 
  row the corresponding
  planetographic latitude.} 
\center
\begin{tabular}{lrrrrr}
\noalign{\smallskip\hrule\smallskip}
                 & S       & C       & E       & R       & total slit \cr
\noalign{\smallskip\hrule\smallskip}
\noalign{location}
$x_{\rm cent}$           & $-9.3''$& $ 0.0''$& $+4.0''$& $+8.0''$     \cr
latitude             & $-65^\circ (^a)$
                                  & $-30^\circ$ 
                                            & $0^\circ$
                                                      & $+30^\circ$   \cr 
\noalign{continuum 5900-6100 \AA , $A_g=0.54$ \hfill}
$I/I_{\rm slit}$ & 0.037   & 0.124   & 0.129   & 0.084   & 1.000      \cr
$f$              &         & 0.68    & 0.71    & 0.46    &            \cr
$Q_r/I$ [\%]     & $+1.40$ & $+0.34$ & $+1.65$ & $-0.10$ & $+0.54$    \cr
$U_r/I$ [\%]     & $+0.24$ & $+0.08$ & $+0.68$ & $-0.17$ & $+0.16$    \cr

\noalign{continuum 8100-8300 \AA , $A_g=0.56$ \hfill}
$I/I_{\rm slit}$ & 0.027   & 0.125   & 0.130   & 0.088   & 1.000      \cr
$f$              &         & 0.71    & 0.74    & 0.50    &           \cr
$Q_r/I$ [\%]     & $+0.40$ & $+0.16$ & $+0.02$ & $-0.21$ & $+0.05$    \cr
$U_r/I$ [\%]     & $+0.15$ & $+0.02$ & $-0.22$ & $-0.16$ & $-0.05$    \cr
\noalign{CH$_4$ 8800-8940 \AA , $A_g=0.07$ \hfill}
$I/I_{\rm slit}$ & 0.045   & 0.047   & 0.102   & 0.301   & 1.000      \cr
$f$              &         & 0.09    & 0.20    & 0.6     &           \cr
$Q_r/I$ [\%]     & $+0.27$ & $+0.19$ & $-0.35$ & $-0.40$ & $-0.18$    \cr
$U_r/I$ [\%]     & $+0.36$ & $-0.04$ & $-0.74$ & $-0.50$ & $-0.38$    \cr
\noalign{\smallskip\hrule\smallskip}
\end{tabular}
\noindent
$a$: southern limb beyond the south pole 
at longitude $+180^\circ$ from the central meridian   
\label{saturntable}
\end{table} 

Our relative $I/I_{\rm slit}$ for Saturn can be converted into reflectivities
$f_\lambda$ using the HST observations from December 2002, published by 
\citet{perezhoyos05}. From the
electronic figure we obtain $f_{6750}=0.72\pm0.01$ 
for the sub-Earth point ``C''. Temporal changes in $f_\lambda$ for Saturn 
can be significant, due to the strong inclination of the planet and the
variable shadowing by the ring system. But the data from December 2002
are useful for us because 
no strong reflectivity changes were noticed between 
Dec. 2002 and Aug. 2003 \citep{perezhoyos05} and between
March 2003 and March 2004 \citep{karkoschka05}. 

Reflectivities for other wavelengths are deduced with 
the global albedo spectrum of \citet{karkoschka98},
according to $f_\lambda = f_{6750}\, A_g(\lambda)/A_g(6750)$, using
$A_g(6750) = 0.57$. The reflectivities for the slit sections 
S, E, and R scale like the $I/I_{\rm slit}$ values (Table \ref{saturntable}).

\subsubsection{Spectropolarimetry for Saturn}

\begin{figure}
\epsscale{0.8}
\plotone{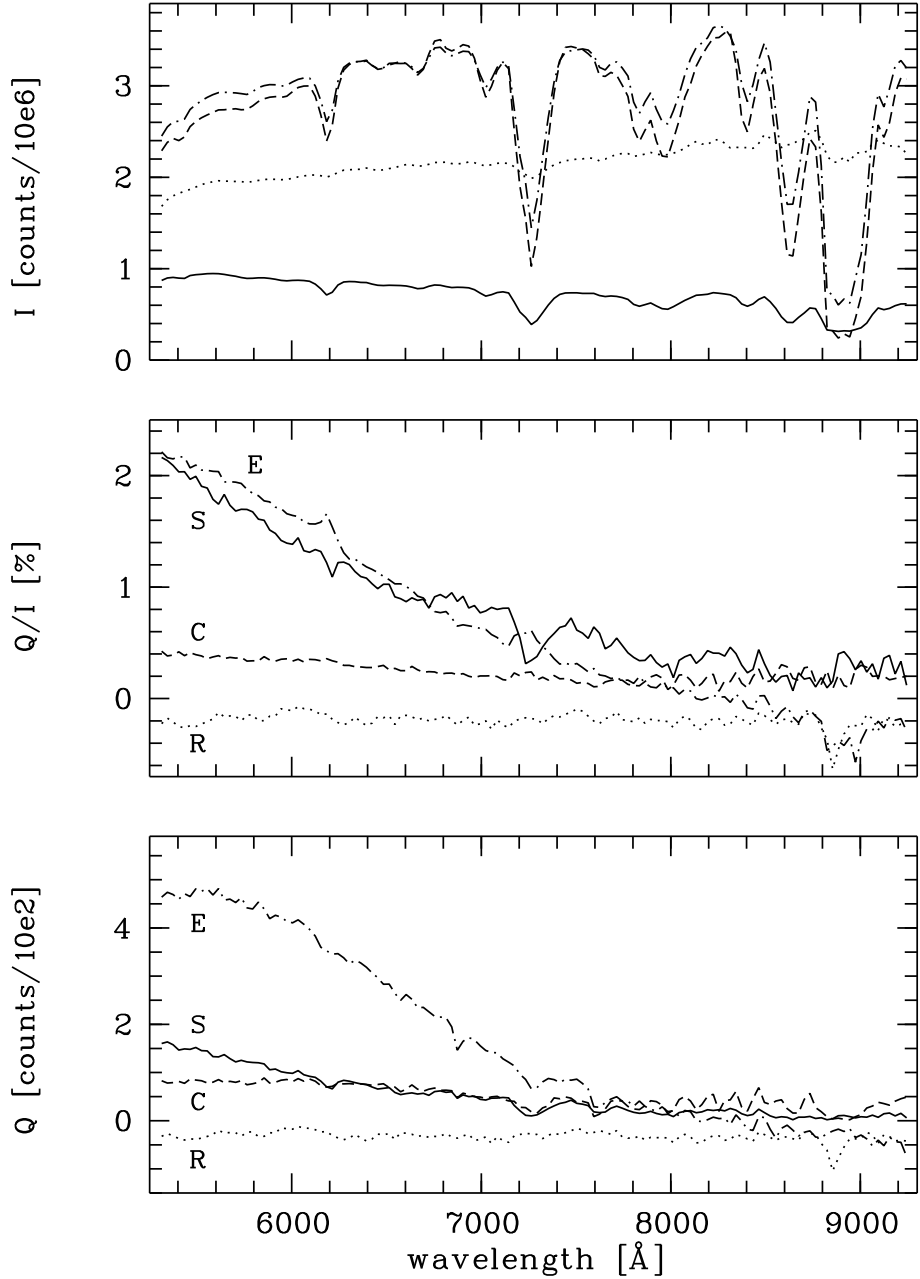}
\caption{Spectropolarimetry for the four regions of Saturn: south pole
  (solid line), center (dashed line), equator
  (dot-dashed line), and ring (dotted line). Top panel: intensity spectra
  $I(\lambda)$ (counts). Middle panel: fractional polarization 
  $Q_r/I(\lambda)$. Positive is parallel to the slit and negative
  is perpendicular. Bottom panel: polarized flux $Q_r(\lambda)$. 
} 
\label{saturnspec}
\end{figure}

The spectropolarimetric signal for the southern limb (S), disk center
(C), equator (E), and 
the ring (R) are plotted in Fig.~\ref{saturnspec}. 
The spectra are averages for 
$2''$ wide spatial regions as indicated in Fig.~\ref{saturncut}. 

The intensity spectra S, C, and E from the planetary 
disk show all the well known
CH$_4$-bands. The ring spectrum R is essentially featureless apart from very
weak features at the wavelength of strong CH$_4$ bands, which can be attributed
to scattered light from the adjacent planetary disk.

A strong fractional polarization ($Q_r/I >1~\%$) is present
at the south pole, ``S'', and equator, ``E'', respectively. 
$Q_r/I$ decreases steeply towards longer
wavelengths from about $Q_r/I\approx +2.2~\%$ at 5300~\AA\ to about 
$+0.2~\%$ at 8000~\AA, and is close to zero for longer wavelengths.
The fast drop in polarization in $Q_r/I$ or in the polarization flux $Q_r$
with wavelength is typical for a 
Rayleigh scattering layer above a weakly or non-polarizing cloud surface
\citep[e.g.][]{buenzli09}. 
 
$Q_r/I$ for the disk center is low, only $+0.4~\%$ at
5300~\AA , and it decreases steadily towards zero. It is not clear, whether the
negative feature in the CH$_4$-band at $\lambda$8870 is real or just a 
spurious effect. As shown in previous studies
\citep[e.g.][]{kemp73,dollfus96} the fractional polarization of the
ring depends only little on wavelength. A weak feature
is visible in the strong CH$_4$-band $\lambda$8870,
which could be due to enhanced contamination from scattered light. 

For the ``E'' and ``S'' regions there are essentially no
narrow spectral features visible
in $Q_r/I$. In particular there is no clear enhancement in  
strong methane bands like $\lambda$7270 or $\lambda$8870, 
as observed for Jupiter (see
Sect.~\ref{Jspecpol}), or Uranus and Neptune \citep{joos07a}.

\section{Investigation of the strongest polarization features}
\label{obsmodels}

Our polarimetry of Jupiter and Saturn has revealed two
surprising polarization features: a very high limb polarization
reaching a maximum of more than 9~\% between 5300~\AA \ and 6500~\AA \
for the South pole of Jupiter, and a strong equatorial polarization for
Saturn. We explore whether these observational results are
compatible with simple scattering models. 

For a detailed characterization of the scattering 
particles in Jupiter and Saturn extensive modelling would be required. 
Unfortunately there exist up to now only few limb polarization 
models -- essentially only the grid 
for Rayleigh scattering atmospheres by \citet{buenzli09} and a few 
previous, mostly analytic results 
as summarized in \citet{schmid06a}. 
Therefore it is not well explored 
yet how the limb polarization depends on the  
scattering phase matrix of the haze particles, the stratification
of the atmosphere, and the optical depth of absorbers. 
Therefore, our model fitting remains ambiguous without
extensive model simulations which are beyond the scope of this paper.

\subsection{Polarization model for the poles of Jupiter}

Model calculations 
for Jupiter were carried out in order to explain the observed 
peak limb polarization of more than 9~\% in R-band and more than 9.5~\% 
in the V-band at the South pole. Considering that the 
seeing degrades this polarization, the maximum limb 
polarization must be well above 10~\%. Rayleigh scattering models 
yield up to 10~\% limb polarization \citep[e.g.][]{schmid06a} 
but only for highly absorbing
atmosphere models, which are not appropriate for the reflected intensity seen
on the poles of Jupiter. 

Detailed scattering models for the polarization of 
Jupiter were presented by \citet{smith84} and \citet{braak02}, but only
for mid and low latitudes. No detailed scattering  models
exist for the polarization at the poles. \citet{smith84} 
made a simple fit to the polarization measured in the red with the
Pioneer spacecraft 
for a Rayleigh scattering layer 
with single-scattering albedo of $\omega=0.983$, optical thickness 
$\tau=0.5$, 
and a surface albedo of $A_S=0.67$. 
However, this Rayleigh scattering model yields a maximum 
limb polarization of only 7.3~\%, or $\approx 6.5~\%$ if the
degradation by the seeing is considered, whereas our measurements 
show a much higher fractional polarization. 

We calculate the polarization along the central meridian 
of Jupiter, considering three zones: the polar S+ and N+ zones, and a
central zone, corresponding roughly to the S$-$, N$-$, and center regions
in Table~\ref{jupitertable}.

For the S+ and N+ zones we use the Monte Carlo multiple scattering
code described in \citet{buenzli09}. The chosen 
atmosphere structure is very similar to the haze model presented by 
\citet{smith84} for low latitudes. They 
determined haze and gas properties for the South Tropical Zone 
and South Equatorial Band with an atmosphere consisting of a top gas 
layer G1, a scattering haze layer H, a lower
gas layer G2, and an optically thick surface layer S at the bottom. 
We fit the polar regions S+ and N+ with this model. 
For the wavelength 6000~\AA \ the
optical depths for the two gas
layers are $\tau_{\rm G1} = 0.011$ and $\tau_{\rm G2}= 0.018$, and the
single scattering albedo is $\omega_{\rm G}=0.976$, calculated
for a methane abundance of 0.18~\%. Thus, the haze layer 
is geometrically thin and located at a pressure level of 290~mbar,
while the opaque surface layer is at 760~mbar. The continuum
polarization at 6000~\AA \ does not depend much on the exact height of the
haze and cloud surface layers and e.g. 40~mbar or 1200~mbar, respectively, 
would not notably change the intensity and polarization. The 
situation is different for the
polarization in the CH$_4$ bands which depends in various ways on the
pressure level of the different layers.
   
The free parameters of our model are: the albedo 
cloud layer $A_{\rm S}$ assumed to be a grey Lambert surface, 
the optical thickness
of the haze layer $\tau_{\rm h}$, and the haze parameters, 
which are single scattering albedo $\omega_{\rm h}$,
single term Henyey-Greenstein asymmetry parameter 
$g_{\rm h}$, and maximum polarization for right angle 
scattering $p_{\rm h}$ \citep[see][]{braak02}. 
The polar model is independent of latitude, but the incidence
and viewing angles produce a latitude dependence 
in the reflected polarization and intensity. 

For the reflected intensity at lower latitudes the same model
is used, but for the fractional
polarization an ``artificial'' constant value of $Q_r/I=-0.7~\%$ is adopted.
The negative polarization is introduced to fit the transition 
between the positively polarized polar zones and the negative 
central zone. The borders $r_{\rm S}$ and $r_{\rm N}$ are 
free parameters which are determined
in the data fitting process. 

In order to describe the smearing of the signal due to
atmospheric seeing and instrumental light scattering, the ``discrete''
three zone model is convolved with a Moffat 
\citep{moffat69} point spread function (PSF).
In the formalism of
\citet{trujillo01} this PSF includes a 
$\beta$-parameter which describes the scattering wing. A small 
$\beta$ implies strong wings, a Gaussian is obtained for 
$\beta \rightarrow \infty$, while atmospheric turbulence theory 
predicts $\beta = 4.76$. For our observations we derive
$\beta=1$ from the residual light outside the nominal limb,
indicating significant scattering in the instrument.  

\begin{figure}
\epsscale{0.8}
\plotone{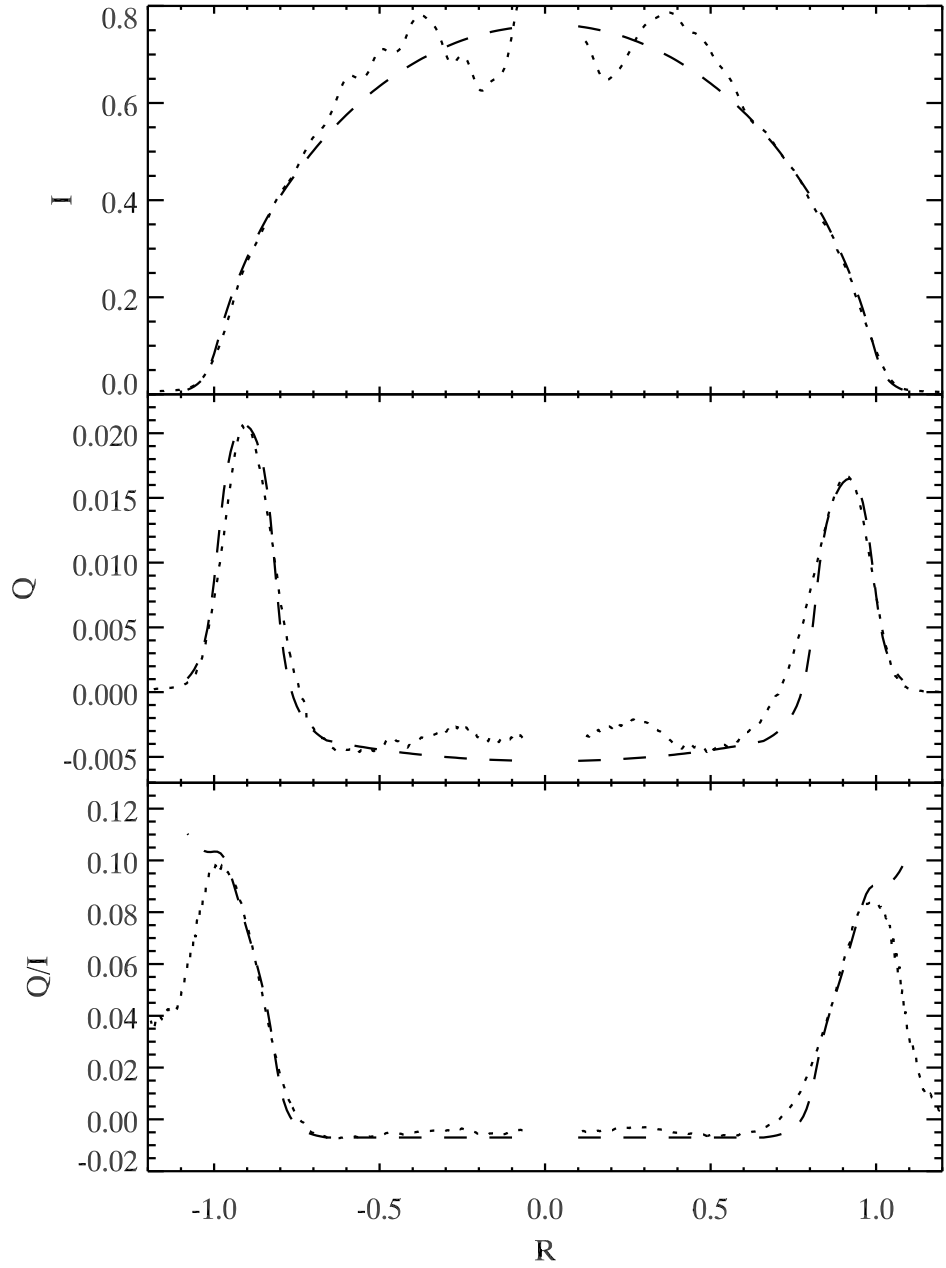}
\caption{Comparison of the model fit (dashed line) with the observed
6000~\AA \ N-S profile (dotted line) of Jupiter.} 
\label{jupitermodel}
\end{figure}

In Fig. \ref{jupitermodel} the observed intensity and polarization 
profiles for the central meridian for 6000~\AA \ are compared with 
the model fit. At the poles 
$|r|>0.8$ the match is satisfactory except for the fractional polarization
outside the limb $|r|>1$, where the statistical errors are large because 
$I\rightarrow 0$. At low latitudes there are some discrepancies 
because we did not try to fit the band structure for the 
intensity, and we adopted a constant value for the fractional polarization.

A good fit for the limb polarization at $\lambda=6000$~\AA\ for 
both poles is obtained for  
strongly polarizing $p_{\rm h}=1$ haze with a low absorption 
$\omega_{\rm h}=0.99$ and asymmetry parameter $g_{\rm h}=0.6$. 
Such scattering parameters are typical for aggregate particles
as proposed to be present in Jupiter and Titan by \citet{west91}. 
Small Mie scatterers with diameters much smaller
than the wavelength of the scattered light have similar scattering
parameters. But the scattering cross section for small spheres 
is much higher in the blue and one would expect a fractional limb
polarization which decreases rapidly with wavelength (as seen
for the pole and the equator of Saturn). The scattering cross section
and therefore the scattering optical depths of the haze layer 
are expected to decrease only slowly
with wavelength for aggregate particles and they can therefore 
explain the rather gentle
decrease in the $Q/I$ and $Q$-spectra towards longer 
wavelengths observed for the poles of Jupiter (see Fig.~\ref{jupspec}).   

The measured limb polarization at the poles for $r=\pm 0.96$ 
is about 9.8~\% in the South and
8.4~\% in the North. The North-South differences can be explained by
different optical depths $\tau_{\rm h}(N+)=0.72$, $\tau_{\rm h}(S+)=1.1$
for the haze layers.

Our model fit includes the smearing due 
to seeing and the modelling indicates that the intrinsic  
limb polarization reaches maxima of about 11.5~\% in the South
and 10.0~\% in the North. Such a high
limb polarization is probably only possible 
with particles having a scattering phase function 
with reduced backscattering, with an asymmetry parameter comparable to 
$g\approx 0.6$ as used in our model. 

Our model fit is not unique. However, many parameters are 
already well constrained. From the parameter space explored 
by us it seems very likely that
the asymmetry parameter lies in the range $0.8< g_{\rm h}< 0.5$ and the
single scattering polarization is $p_{\rm h} > 0.8$.

The locations of the transitions $r_{\rm S}$
and $r_{\rm N}$ between the highly 
polarized poles and the negatively polarized central zone are at
$r=\pm 0.82\pm0.01$, identical for both hemispheres. 
This correspond to a planetographic latitude of 
$\pm 58^\circ\pm 3^\circ$. In the South the 
transition between the S+ and the central region is 
compatible with an unresolved discontinuity. In the North the transition is
more gradual, with a weak wing of positive polarization towards 
lower latitude. 
This is in qualitative agreement 
with Fig.~14 in \citet{smith84},
who measured the latitudinal polarization dependence 
with Pioneer for large phase angles (82$^\circ$ and $98^\circ$). 
Their red filter data show a steep increase in the polarization
in the South from about 10~\% at a latitude of $-55^\circ$ to about 33~\% at
$-65^\circ$ and a more gradual 
increase from 13~\% at $+55^\circ$ to about 25~\% at $+65^\circ$ in the
North.  

The surface albedo $A_S=0.75$  
is also quite well constrained, since a high value $A_S>0.9$ 
would significantly reduce the resulting fractional polarization 
and a low value $A_S<0.6$ would underpredict the reflectivity at the poles. 
It also seems quite safe to explain the North-South asymmetry in the polar
polarization with a difference in the optical thickness of the polarizing
haze layer. 
   
Finally, it is interesting to note that the haze model for low latitudes 
with $\omega_h=0.95 $, $p_{\rm h}=0.9$ and $g_h=0.75$ 
from \citet{smith84}, which was also adopted by 
\citet{braak02}, cannot fit the
polarization at the poles. This points to  
distinct differences between the haze particles at the poles 
and at lower latitudes.

\subsection{The transient polarization feature at the equator of Saturn}

For Saturn we investigate the nature of the surprisingly strong transient 
polarization feature at the equator $Q_r/I(6000~{\rm \AA})=+1.65~\%$
(Table \ref{saturntable}), present in our 2003 (and 2002)
data. The transient nature of the feature could be
related to the high inclination of $\approx 26^\circ$ 
of the Saturn system during our observations. In the following
we explore possible explanations for this feature.

At exact opposition no polarization signal is expected from
the center of the planetary disk for
symmetry reasons. Our spectropolarimetric 
observations were taken at phase angle 
$\alpha=3.7^\circ$. The equatorial polarization cannot be
ascribed to a phase effect because for this phase one expects only
a negligible positive polarization for small, positive 
polarizing scattering particles,
or a small $Q/I\approx -0.2~\%$ negative  
polarization for e.g. large negatively polarizing scattering particles.   

Higher order scatterings produce a radial limb 
polarization, but this initially increases only slowly when going from
the center of the disk towards the limb. A semi-infinite, conservative 
($\omega=1$) Rayleigh scattering atmosphere produces a fractional limb 
polarization of only $Q_r/I=+1.3~\%$ at 
$r/r_{\rm limb}\approx 0.5$ or $\theta\approx 30^\circ$
\citep[e.g.][]{buenzli09,schmid06a} which is
roughly the location of the equator in our observations. 
A slightly higher limb polarization is possible with
a stratospheric haze model similar to the one determined by
\citet{karkoschka05} consisting of a 
relative thick layer with high albedo, highly polarizing, 
and forward scattering particles 

However, the interpretation of the strong equatorial polarization in
Saturn as higher order scattering 
or ``limb polarization'' effect is in contradiction
with the observations. Limb polarization would produce a $Q_r/I$
signal along the entire equatorial band with maxima
above 5~\% at the equatorial limbs. This is not observed.
In addition we would
expect a high polarization of at least $p>20$~\% at large phase
angles which is also not visible in the contemporaneous Cassini
polarization image of \citet{west09}.

An alternative explanation could be strong light scattering from the 
inclined ring system. But this can be rejected with 
an order of magnitude estimate.
The solid angle subtended by the illuminated ring system as seen by the
equatorial region is small and the forward scattering of light by the ring is 
inefficient because the ring is made of large ($>1$~mm), 
back-scattering bodies. For example \citet{dones93} show
that the phase function for forward scattering by the ring is only 
$I(\alpha)<0.1\,I(0)$ for $\alpha>135^\circ$. Therefore the
irradiation with light scattered from the rings
is well below 1~\% of the direct irradiation from the sun and it
is impossible that this low level of indirect light produces a significant
fractional polarization at the equator.   

Which other effect could explain the transient equatorial polarization
of Saturn if phase effects, limb polarization, and ring reflection can be
discarded? 
The unknown effect seems to be restricted to the equatorial band and 
photon incidence and emergence 
angles of $\approx 25^\circ$ for the incoming and back-scattered radiation. 
Such a localized feature could perhaps be formed by a higher order scattering 
effect in an optically thick, but geometrically thin cloud or aerosol 
layer. We are not aware of investigations of such scattering geometries. 
However, we may hope 
that a detailed analysis of the Cassini polarimetry presented 
in \citet{west09} clarifies the case.

\section{Summary, discussion, and conclusions}

This work presents ground-based spectropolarimetry and 
imaging polarimetry of Jupiter and Saturn. The new data 
are of unprecedented quality because 
modern instruments with CCD detectors were used. This type of 
data opens up many new avenues of polarimetric investigation for
planets, because the polarization signal can be quantified much more
accurately from imaging polarimetry and spectropolarimetry, when 
compared to previous aperture polarimetry. 
Well calibrated data with quantified seeing effects allow
for detailed comparisons with model calculations of 
the scattering layers in Jupiter and Saturn.  
\smallskip

\noindent
There are a series of important results in this work: 

We present for the first time spectropolarimetry
for the strongly polarized poles of Jupiter. 
The polarization shows an overall decrease with 
wavelength and a rich spectral structure with 
enhanced fractional polarization in strong methane absorptions. In the
polarization flux spectrum the absorption bands are still visible 
as absorptions, although with a lower equivalent width than in the
intensity spectrum. The enhancement in fractional
polarization $Q/I$ in the methane bands can be explained 
by a highly polarizing stratospheric haze layer 
above an atmosphere where multiple scatterings (or the 
reflection of unpolarized light) are suppressed in 
the methane absorptions.  

The fractional polarization at lower latitudes of Jupiter is small and 
slightly negative with an enhanced polarization signal in strong
methane absorption. 

From the polarization profiles along the central
meridian of Jupiter we can derive a latitude of $\pm 58^\circ$
for the transition between the polar region 
with strong positive polarization 
and lower latitudes with a negative polarization. Our results agree very well
with the polarization measurements from the Pioneer missions 
\citep{smith84}. Stratospheric haze is responsible for the
strong polar polarization and the sharp transition is pointing to a  
well defined border in the stratospheric circulation.

For the poles of Jupiter we measure in the V-band a seeing limited 
peak limb polarization of $+9$ to $+10~\%$. This indicates a 
resolution-corrected peak polarization of about $+11.5~\%$. 
All previous studies reported
lower values ($p\approx 6-8$~\%) for the maximum polarization at the poles
of Jupiter. This difference can be explained by the better
spatial resolution (seeing $\approx 1''$) of our data and the appropriate 
seeing correction. 

We derive the 
polarization flux of the entire positively polarized 
polar hoods (called $S+$ and $N+$ in this work). 
This approach should allow an easy comparison of our data
with previous and future observations for long term studies
which are interesting for investigations of the haze 
production, destruction and transport 
in the polar stratosphere of Jupiter. Such long term polarization
changes were reported e.g. by \citet{starodubtseva02}
and \citet{starodubtseva09}.

We found no limb polarization models with a
limb polarization higher than 10~\% in the literature. For this
reason we carried out exploratory calculations. They indicate that 
a polar haze layer consisting of forward scattering and highly 
polarizing aggregate particles as proposed by \citet{west91} 
is compatible with our observations. 
\smallskip

The polarization of Saturn is lower than that of Jupiter. 
Our observations of 2002 and
2003 revealed at 6000~\AA\ a positive feature (about $1.4~\%$) at the 
South pole and a stronger ($>1.6~\%$) transient polarization feature 
near the equator. 

Spectropolarimetrically, Saturn shows hardly any enhancements in 
the fractional polarization at wavelengths of strong methane bands 
in strong contrast to the case of Jupiter. This indicates that the 
polarizing particles in Saturn are not located in a discrete layer,
but well distributed in the 
scattering atmosphere. Absorptions within the polarizing scattering layer
reduce the reflected intensity $I$ and polarization flux $Q$ simultaneously
and causes therefore no or only weak features in the fractional polarization
spectrum $Q/I$. 

At the South pole and the equator the
polarization decreases rapidly with wavelength, and essentially no
polarization $p < 0.4~\%$ is present for wavelengths above $7500$~\AA.   
The fast decrease of the limb polarization with wavelength
is indicative for small scattering particles ($< 100$~nm) as  
proposed by \citet{karkoschka05} based on 
HST imaging. 

The strong equatorial polarization feature in Saturn is 
remarkable since it was
not present in earlier measurements. During our
observations the inclination of Saturn was high and 
therefore the effect could be seasonal. 
We investigated various possibilities to explain the equatorial  
polarization but could not find a reasonable solution. Despite this
we are certain that the observed fractional polarization 
of about $p=1.6~\%$ in polar direction is real. It seems
interesting to investigate this feature further with other data or
model simulations. 
\smallskip

An important conclusion from this paper is that 
modern polarimetric measurements of Jupiter and Saturn from 
the ground can provide accurate quantitative results 
which constrain strongly the scattering properties of 
the atmospheres. This type of investigations certainly 
did not receive sufficient attention in the last decades, 
and more observations, e.g. for other wavelengths, and
a lot of limb polarization modelling still needs to be done. 
A first step with model simulations was done by \citet{buenzli09} 
who present an extensive grid of model calculations for 
Rayleigh scattering atmospheres. Further models
ought now to be calculated which should investigate the 
dependence of the limb polarization signal on the
scattering phase matrix of different populations of haze particles. 
In additions one needs to take into account the spectropolarimetric 
structure in the methane bands in order to derive the 
vertical stratification of the scattering layers. 

Scattering layers, reflecting the solar light, are an intriguing aspect
of solar system planets. They affect the radiative
transfer in these objects. For the investigation of the reflected
light from extra-solar planets a comprehensive understanding of 
the physics of the high altitude haze layers is very important.
For this reason it is essential to carry out detailed investigations of
the reflecting layers in solar system planets. Investigations 
based on modern polarimetric observations, as presented in this work for
Jupiter and Saturn, are therefore very valuable for progress in 
this direction.

\noindent 
{\sl Acknowledgements:} 
We are indebted to the ESO La Silla support team at the 3.6m telescope
who were most helpful with our very special EFOSC2 instrument setup. We are
particularly grateful to Oliver Hainaut. 
We thank Harry Nussbaumer for carefully reading the manuscript. We 
also acknowledge the many useful comments from the referees. 
This work was supported by a grant from the Swiss National Science
Foundation.

{}

\clearpage


\begin{thebibliography}{}
\bibitem[The Astronomical Almanach(2003)]{almanach03}
        The Astronomical Almanach, 2003, US Gov. Printing Office Washington,
        The Stationary Office, London
\bibitem[Braak et al.(2002)]{braak02}
        Braak, C.\,J., de Haan, J.\,F., Hovenier, J.\,W., Travis, L.\,D.
        2002.
        Galileo Photopolarimetry of Jupiter at 678.5 nm. 
        Icarus 157, 401-418.
\bibitem[Buenzli and Schmid(2009)]{buenzli09}
        Buenzli, E., Schmid, H.M., 2009. 
        A grid of polarization models for Rayleigh scattering 
        planetary atmospheres.
        A\&A 504, 259-276
\bibitem[Carlson and Lutz(1989)]{carlson89}
        Carlson, B.E., Lutz, B.L., 1989. Spatial and temporal variations
        in the atmosphere of Jupiter: polarimetric and photometric 
        constraints. 
        In: Belton, M.J.S., West, R.A., Rahe, J., Proc. of Workshop on
        Time-variable phenomena in the Jovian System, 
        NASA special publication series, NASA-SP-494, pp. 289-298 
\bibitem[Chanover et al.(1996)]{chanover96}
        Chanover, N.J., Kuehn, D.M., Banfield, D., Momary, T., Beebe,
        R.F., Baines, K.H., Nicholson, P.D., Simon, A.A., Murrell, A.S., 1996.
        Absolute Reflectivity Spectra of Jupiter: 0.25-3.5 Micrometers.
        Icarus 121, 351-360.
\bibitem[Cochran et al.(1981)]{cochran81}
        Cochran, A.L., Trafton, L.M., Cochran, W.D., 
        Barker, E.S., 1981. 
        Spectrometry of Jupiter at selected locations on the disk during 
        the 1979 apparition.
        AJ 86, 1101-1107.
\bibitem[Coffeen(1979)]{coffeen79}
        Coffeen, D.L., 1979.
        Polarization and scattering characteristics in the atmospheres of 
        Earth, Venus, and Jupiter. 
        J. Opt. Soc. Am. 69, 1091 
\bibitem[Dollfus(1957)]{dollfus57}
        Dollfus, A., 1957. 
        \'Etude des plan\`etes par la polarisation de leur lumi\`ere.
        Suppl. Ann. Astrophys. 4, 3-114,
\bibitem[Dollfus(1979)]{dollfus79}
        Dollfus, A., 1979.
        Optical reflectance polarimetry of Saturn globe and rings. 
        I - Measurements on B ring. 
        Icarus 37, 404-419.
\bibitem[Dollfus(1996)]{dollfus96}
        Dollfus, A., 1996. 
        Saturn's Rings: Optical Reflectance Polarimetry.
        Icarus 124, 237-261.
\bibitem[Dones et al.(1993)]{dones93}
        Dones, L., Cuzzi, J.~N., Showalter, M.~R.\ 1993.\ 
        Voyager Photometry of Saturn's A Ring.\ Icarus 105, 184-215. 
\bibitem[Gehrels et al.(1969)]{gehrels69}
        Gehrels, T., Herman, B.\,M., Owen, T., 1969. 
        Wavelength Dependence of Polarization. XIV. Atmosphere of Jupiter.
        Astron. J. 74, 190-199.
\bibitem[Gisler et al.(2003)]{gisler03}
        Gisler, D., Schmid, H.M., 2003. 
        Non-solar Applications with the ZIMPOL Polarimeter.
        In: Trujillo Bueno, J., Sanchez Almeida, J. (Eds.), Conference on 
        Solar Polarization, September 30 - October 4 2002, Tenerife, Spain,
        ASP Conf. Ser. 307, 58-61.     
\bibitem[Gisler et al.(2004)]{gisler04}
        Gisler, D., Schmid, H.M., Thalmann, C., Povel, H.P., Stenflo, J.O., 
        Joos, F., et al., 2004. 
        CHEOPS/ZIMPOL: a VLT instrument study for the polarimetric 
        search of scattered light from extrasolar planets.
        In: Moorwood, A.F.M., Iye, M. (Eds.), SPIE conference on 
        Ground-based instrumentation for Astronomy, 
        Proc. SPIE Vol. 5492, 463-474. 
\bibitem[Hall and Riley(1968)]{hall68}
        Hall, J.\,S., Riley, L.\,A., 1968. 
        Photoelectric observations of Mars and Jupiter with a 
        scanning polarimeter.
        Lowell Obs. Bull. 7, 83-92.
\bibitem[Hall and Riley(1974)]{hall74}
        Hall, J.\,S., Riley, L.\,A., 1974. 
        A photometric study of Saturn and its rings.
        Icarus 23, 144-156.
\bibitem[Hall and Riley(1976)]{hall76}
        Hall, J.\,S., Riley, L.\,A., 1976. 
        A polarimetric search for fine structure on Jupiter's disk.
        Icarus 29, 231-234.
\bibitem[Hough et al.(2006)]{hough06}
        Hough, J.H., Lucas, P.W., Bailey, J.A., Tamura, M., 
        Hirst, E., Harrison, D., 
        Bartholomew-Biggs, M., 2006. 
        PlanetPol: A Very High Sensitivity Polarimeter.
        PASP 118, 1302-1318.
\bibitem[Johnson et al.(1980)]{johnson80}
        Johnson P.E., Kemp J.C., King R., Parker T.E., Barbour M.S.,
        1980. 
        New results from optical polarimetry of Saturn's rings.
        Nature 28, 146-149. 
\bibitem[Joos and Schmid(2007a)]{joos07a}
        Joos, F., Schmid, H.M., 2007a. 
        Limb polarization of Uranus and Neptune. 
        II. Spectropolarimetric observations.
        A\&A 463, 1201-1210.
\bibitem[Joos and Schmid(2007b)]{joos07b}
        Joos, F., Schmid, H.M., 2007b. 
        Polarimetry of Solar System Gaseous Planets.
        The Messenger, 130, 27-31.
\bibitem[Karkoschka(1998)]{karkoschka98}
        Karkoschka, E., 1998.
        Methane, Ammonia, and Temperature Measurements of the 
        Jovian Planets and Titan from CCD-Spectrophotometry. 
        Icarus 133, 134-146
\bibitem[Karkoschka and Tomasko(2005)]{karkoschka05}
        Karkoschka, E., Tomasko, M., 2005.
        Saturn's vertical and latitudinal cloud structure 1991-2004 
        from HST imaging in 30 filters.
        Icarus 179, 195-221.
\bibitem[Kattawar and Adams(1971)]{kattawar71}
        Kattawar, G.W., Adams, C.N., 1971. 
        Flux and Polarization Reflected from a Rayleigh-Scattering 
        Planetary Atmosphere.
        ApJ 167, 183-192.
\bibitem[Kemp et al.(1971)]{kemp71}
        Kemp J.C., Wolstencroft R.D., Swedlund J.B., 1971. 
        Circular Polarization: Jupiter and Other Planets.
        Nature 232, 165-168.
\bibitem[Kemp and Murphy(1973)]{kemp73}
        Kemp, J.C., Murphy, R.E., 1973. 
        The Linear Polarization and Transparency of Saturn's Rings.
        ApJ 186, 679-686.
\bibitem[Leroy(2000)]{leroy00}
        Leroy, J.L., 2000. 
        Polarization of light and astronomical observations. 
        Chapts.~5 and 6, Gordon \& Breach.
\bibitem[Lyot(1929)]{lyot29}
        Lyot, B., 1929. Recherches sur la polarisation de la lumi\`ere des 
        plan\`tes et de quelques substances terrestres. 
        Ann. Observ. Meudon, VIII, 1-144. English translation, NASA TT F-187.
\bibitem[Moffat(1969)]{moffat69}
        Moffat, A.F.J., 1969, A Theoretical Investigation of Focal 
        Stellar Images in the Photographic Emulsion and Application 
        to Photographic Photometry. A\&A 3, 455-461
\bibitem[Moreno et al.(1991)]{moreno91}
        Moreno, F., Molina, A., Lara, L.M., 1991. 
        Charge-coupled device spectral images of spatially resolved 
        regions of Jupiter in the 6190 and 8900 \AA\ methane and 
        6450 \AA\ ammonia bands during the 1989 opposition.
        Journal of Geophys. Res. 96, 14119-14127.
\bibitem[Morozhenko(1973)]{morozhenko73}
        Morozhenko, A.\,V., 1973. 
        Polarimetric Observations of the Giant Planets. III. Jupiter.
        Soviet Astronomy, 17, 105-107.
\bibitem[Ortiz et al.(1995)]{ortiz95}
        Ortiz, J.L., Moreno, F., Molina, A., 1995. 
        Saturn 1991-1993: Reflectivities and limb-darkening 
        coefficients at methane bands and nearby continua -- temporal changes.
        Icarus 117, 328-344.
\bibitem[Perez-Hoyos et al.(2005)]{perezhoyos05}
        Perez-Hoyos, S., Sanchez-Lavega, A., French, R.G., Rojas,
        J.F., 2005. 
        Saturn's cloud structure and temporal evolution from ten 
        years of Hubble Space Telescope images (1994-2003).
        Icarus 176, 155-174.
\bibitem[Povel(1995)]{povel95}
        Povel, H., 1995. 
        Imaging Stokes polarimetry with piezoelastic modulators and 
        charge-coupled-device image sensors.
        Optical Engineering 34, 1870-1878.
\bibitem[Schmid et al.(2006a)]{schmid06a}
        Schmid, H.\,M., Joos, F., Tschan, D., 2006.  
        Limb polarization of Uranus and Neptune. I. Imaging polarimetry 
        and comparison with analytic models.
        A\&A 452, 657-668. 
\bibitem[Schmid et al.(2006b)]{schmid06b}
        Schmid, H.M., Beuzit, J.-L., Feldt, M., Gisler, D., Gratton, R.,
        Henning, Th., Joos, F., Kasper, M., Lenzen, R., Mouillet, D., 
        Moutou, C., Quirrenbach, A., Stam, D. M., Thalmann, C., 
        Tinbergen, J., V\'erinaud, C., Waters, R., Wolstencroft, R., 2006.  
        Search and investigation of extra-solar planets with polarimetry.
        In: Aime, C., Vakili, F. (Eds.), IAU colloquium 200 on 
        Direct Imaging of Exoplanets, Cambridge University Press, pp. 165-170. 
\bibitem[Seager et al.(2000)]{seager00}
        Seager, S., Whitney, B.A., Sasselov, D.D., 2000. 
        Photometric Light Curves and Polarization of Close-in 
        Extrasolar Giant Planets.
        ApJ 540, 504-520. 
\bibitem[Shalygina et al.(2008)]{shalygina08}
        Shalygina, O.S., Korokhin, V.V., Starukhina, L.V., 
        Shalygin, E.V., Marchenko, G.P., Yelikodsky, Y.I., 
        Starodubtseva, O.M., Akimov, L.A., 2008.
        The north-south asymmetry of polarization of Jupiter: 
        The causes of seasonal variations.
        Solar system research 42, 8-17.
\bibitem[Smith and Wolstencroft(1983)]{smith83}
         Smith, R.J., Wolstencroft, R.D., 1983. 
         High precision spectropolarimetry of stars and planets. 
         II Spectropolarimetry of Jupiter and Saturn.
         MNRAS 205, 39-55.
\bibitem[Smith and Tomasko(1984)]{smith84}
         Smith, P.H., Tomasko, M.G., 1984. 
         Photometry and polarimetry of Jupiter at large phase angles. 
         II - Polarimetry of the south tropical zone, south equatorial belt, 
         and the polar regions from the Pioneer 10 and 11 missions.
         Icarus 58, 35-73.
\bibitem[Stam et al.(2004)]{stam04}
        Stam, D.M., Howenier, J.W., Waters, L.B.F.M., 2004. 
        Using polarimetry to detect and characterize Jupiter-like 
        extrasolar planets.
        A\&A 428, 663-672.
\bibitem[Starodubtseva et al.(2002)]{starodubtseva02} 
        Starodubtseva, O.~M., Akimov, L.~A., Korokhin, V.~V.\ 2002.\ 
        Seasonal Variations in the North-South Asymmetry of Polarized 
        Light of Jupiter.\ Icarus 157, 419-425. 
\bibitem[Starodubtseva(2009)]{starodubtseva09} 
        Starodubtseva, O.~M.\ 2009.\ Polarization of Jupiter: 
        Semiannual variations in the north-south asymmetry.\ 
        Solar System Research 43, 277-284. 
\bibitem[Swedlund et al.(1972)]{swedlund72}
        Swedlund, J.B., Kemp J.C., Wolstencroft R.D., 1972. 
        Circular Polarization of Saturn.
        ApJ 178, 257-266.
\bibitem[Tomasko and Doose(1984)]{tomasko84}
        Tomasko M.G., Doose L.R., 1984. 
        Polarimetry and photometry of Saturn from Pioneer 11 
        observations and constraints on the distribution and 
        properties of cloud and aerosol particles.
        Icarus 58, 1-34.
\bibitem[Tinbergen and Rutten(1992)]{tinbergen92}
        Tinbergen, J., Rutten, R. 1992. 
        Measuring polarisation with ISIS, 
        Users' manual, The Isaac Newton Group of Telescopes, 
        (http://www.ing.iac.es/)

\bibitem[Trujillo et al.(2001)]{trujillo01}
        Trujillo, I., Aguerri, J.A.L., Cepa, J., Guti\'errez, C.M., 2001.
        The effects of seeing on S\'ersic profiles - II. The Moffat PSF. 
        MNRAS 328, 977-985.

\bibitem[van de Hulst(1980)]{vandehulst80}
        van de Hulst, H.\,C. 1980, Multiple Light Scattering, 2nd Volume, 
        Academic Press, New York

\bibitem[West(1979)]{west79}
       West, R.A., 1979. 
       Spatially resolved methane band photometry of Jupiter. 
       I - Absolute reflectivity and center-to-limb variations in the 6190-, 
       7250-, and 8900-\AA\ bands. II - Analysis of the south equatorial 
       belt and south tropical zone reflectivity.
       Icarus, 38, 12-53.

\bibitem[West et al.(1983)]{west83}
       West, R.A., Sato, M., Hart, H., Lane, A.L., Hord, C.W.,
       Simmons, K.E., Esposito, L.W., Coffeen, D.L., Pomphrey R.B.
       1983. 
       Photometry and polarimetry of Saturn at 2640 and 7500 \AA.
       Journal Geophys. Research, 88, 8679-8697.

\bibitem[West and Smith(1991)]{west91}
       West, R.~A., Smith, P.~H.\ 1991.\ Evidence for aggregate 
       particles in the atmospheres of Titan and Jupiter.\ Icarus 90, 330-333. 

\bibitem[West et al.(2009)]{west09}
       West, R.~A., Baines, K.~H., Karkoschka, E., S{\'a}nchez-Lavega, A.
       \ 2009.\ Clouds and Aerosols in Saturn's Atmosphere.\ Saturn 
       from Cassini-Huygens 161-179. 

\bibitem[Wolstencroft and Smith(1978)]{wolstencroft78}
       Wolstencroft, R.\,D. \& Smith, R.\,J., 1979.
       Spectropolarimetry of the methane and ammonia bands of Jupiter 
       between 6800 and 8200 \AA. 
       Icarus 38, 155-165.
      

\end{thebibliography}
\end{document}